# New Formalism for Modelling Flowing Plasmas in a Magnetic Field

Subhasish Bag[1,2], Ashish Ganguli[3], Vikrant Saxena[1], Ramesh Narayanan[3], Debaprasad Sahu[3]

[1]Department of Physics, Indian Institute of Technology Delhi, New Delhi, India
[2]Asia Pacific Center for Theoretical Physics, Pohang, Gyeongbuk, Republic of Korea
[3]Department of Energy Science and Engineering, Indian Institute of Technology Delhi, New Delhi, India
Corresponding author: vsaxena@physics.iitd.ac.in

**Abstract**

A new formalism for modelling a flowing plasma in a magnetic field, having potential for application in plasma thrusters, materials processing, space plasmas, etc., is developed. In this framework, the plasma expands downstream from the source region into an expansion chamber along an axisymmetric magnetic field. For plasma flowing along a magnetic field, the ion velocity parallel to the magnetic field lines is much greater than the perpendicular velocity. This characteristic permits a unique ordering of the relevant flow variables when the flow equations are transformed into a *magnetic coordinate system* (MCS), in which the coordinate axes are parallel and perpendicular to the field lines. The ordering of the flow variables in the MCS further simplifies the flow equations by allowing them to be split into sets of reduced equations. As a specific implementation, the proposed formalism was validated by using data (for boundary conditions) from a small volume plasma system experiment (Ganguli *et al* 2016 *Plasma Sources Sci. Technol*. **25** 025026). The model shows favourable agreement with the experiments and is used to predict various physical quantities relevant for applications. The corresponding physical implications are discussed in detail.

## 1. Introduction

The study of flowing plasma in a diverging magnetic field is important for modeling plasma thrusters [1-21] and other plasma processing applications [16, 22-34]. In these applications, plasma is allowed to expand downstream from the source region into the expansion chamber. In this context, different experiments have been conducted and analysed [1-28]. In the thruster experiments [1, 4, 12, 46] plasma flow has been characterised in the expansion chamber with a view to determine the thrust imparted by the ions. An external magnetic field is used to produce and guide the expanding plasma. Flowing plasmas also form an essential part of the different plasma-based tools used extensively in semiconductor processing [26], which include applications like surface coating, etching, etc. [16, 17, 27, 28].

Investigations are also being undertaken for better understanding and control of the different technologies and to upgrade or improve them whenever possible for superior end results. For plasma thrusters, experiments are conducted in large expansion chambers (for thrust and beam quality validation) before they are used for actual space applications. Likewise, extensive trial runs need to be conducted before any plasma source can be adopted for processing applications. In this context, the type of plasma source used for the applications is of vital importance since each source has its unique features and one must choose a source that suits the requirements of the specific application. Typically, one uses plasma sources like the inductively coupled plasma (ICP) sources [35, 36], helicon plasma sources (HPS) [ 37-40], electron cyclotron resonance (ECR) sources [36, 41-43], ion cyclotron resonance (ICR) sources [44], etc.

Several of the plasma sources employ magnetic fields for plasma production and plasma extraction. In general, the plasma source is attached to an expansion chamber (EC) where the properties of the downstream plasma are investigated. However, the magnetic field of the source also penetrates the expansion chamber so that the plasma flows into the expansion chamber along these field lines. It turns out that in the different plasma-based applications the magnetic field in the expansion chamber plays an important role in increasing the efficiency of processes and aiding in the optimisation of system performance. For example, in plasma-enhanced chemical vapour deposition (PECVD), magnetised plasmas enhance film uniformity and adhesion by regulating ion trajectories, allowing precise thin-film coatings. Again, in the case of reactive ion etching (RIE) processes, the magnetic field influences the ionisation processes and helps direct ion bombardment for highly anisotropic implantation with minimal damage. Thus, for analysing such applications one must investigate properties of flowing plasmas in magnetic fields.

The electron cyclotron resonance (ECR) mechanism method used for plasma production uses microwaves in the presence of a magnetic field such that the fundamental ECR condition is met. This results in efficient coupling of the microwave power to the electrons and very high-density plasmas may be obtained at fairly low pressures. The Plasma Lab at IIT Delhi has been investigating ECR plasma sources for various applications (for thrusters as well as processing). In particular, experiments using a Compact ECR Plasma Source (CEPS) [45, 46] have been undertaken extensively. The CEPS is a patented device of Plasma Lab, IIT Delhi [Indian Patent No # 301583, Patentee: IIT Delhi]. It comprises a microwave coupler, an impedance matching unit, a mode transformer, and a Plasma Source Section (PSS), all integrated into a single unit. A set of NdFeB ring magnets placed coaxially on the PSS provides the magnetic field required for ECR plasma generation. The device has an overall length, $\simeq$ 60 cm and weighs $\simeq$ 14 kg (including the magnets) [46, 47]. The lab has conducted numerous experiments using the CEPS as the primary plasma producing device by attaching it to chambers of different sizes to test its efficacy for plasma production and generation of intense, high energy ion beams applications in plasma processing as well as plasma thrusters [48-50].

In the experiments, the magnetic field of the CEPS penetrates the expansion chamber (EC) giving rise to diverging field lines in the EC. Plasma from the CEPS flows along these field lines into the EC. Since such conditions are generic to typical plasma-based applications, it was felt that the understanding, utility, and efficiency of such applications would be greatly enhanced if a suitable model for a plasma flowing along an expanding magnetic field could be developed. It may be noted however that it would be very difficult, at least initially, to develop a fully self-consistent model that includes the plasma source because of difficulties in modelling plasma–rf / microwave interactions in complex magnetic field topologies inside the source [45-47]. However, use of the appropriate flow equations to analyse the plasma in the EC alone, would not only render the problem much more tractable, but would also provide useful data about the different plasma parameters that influence and control the plasma processes in the various applications. The present work develops such a model. For ease of computation, the development and validation of the model was based on the smallest expansion chamber under consideration. Thus, the calculations for the model were initialized using data from the Small Volume Plasma System (SVPS: Length ~ 37 cm, Dia. ~ 15 cm) [45-48].
The magnetic field of the CEPS (produced by large, ring magnets of different sizes) was determined using COMSOL and NISA software and verified experimentally at various locations, both inside the magnets and outside. Thus, *numerical values of the field were available within the plasma source section and throughout the chamber*. The magnetic field is azimuthally symmetric, which simplifies the system geometry.

The flow problem requires one to solve for the plasma density $n$, electron temperature $T_e$ and the three velocity components of the ion fluid throughout the expansion chamber. To maintain quasineutrality, the electron flow velocity must equal the ion flow velocity, which is small compared to the electron thermal speeds. To solve the problem one needs to specify suitable boundary conditions. In the present case the boundary conditions comprise values of the flow variables and / or their gradients, in addition to the plasma potential, etc. These must be obtained from the experiments along suitable boundaries of the system. For an azimuthally symmetric magnetic field the flow problem becomes two dimensional ($r$, $z$ coordinates, only) and permits one to consider only *one-half section of the chamber* (from its axis to the side wall), *with the axis becoming* a *boundary* along which all variables may be obtained by measurement.

Boundary conditions play a crucial role in determining the solvability of the flow equations and determining the nature of the solutions. As stated above, experiments designed to investigate the different plasma-based applications may be used to generate data along suitable boundaries of the system for initialising the flow equations. With the boundary conditions known, the main issue that arises is to solve the six (or more), coupled, nonlinear flow equations, involving a magnetic field. In general, such equations are difficult to solve, requiring special techniques and are particularly time consuming with considerable loss of numerical accuracy, when these are solved for large-sized expansion chambers. Thus, procedures that help simplify and reduce the problem complexity without compromising greatly on accuracy are highly welcome.

It turns out that for *plasma flowing along a magnetic field, the flow velocity along the field lines is much greater than the velocities perpendicular to the field*. The latter ordering of the velocities arises naturally in a *magnetic field-based coordinate system* (MCS) where the grid lines are curved and are aligned along the magnetic field or are perpendicular to it. Such an MCS can be set up if a detailed knowledge of the magnetic field in either analytical or numerical form is available. As already noted earlier, a detailed mapping of the CEPS magnetic field was available throughout the chamber as numerical data.

*Using the MCS allows one to partition the variables of the flow problem into two sets*: a *lowest order set* comprising the appropriate variables together with their gradients *along* the field lines and a *second, higher order set*, comprising rest of the variables. This ordering of the variables aids further *the splitting of each equation of the model into a pair of simpler equations: one equation resulting from the lowest order terms and the second, from the higher order terms*. In fact, this splitting of the original flow equations into simpler equations is a crucial step that allows considerable simplification of the problem and helps one to bypass singularities that appear during the integrations. It is this approach that has been adopted in the present work.

Section 2 begins with a brief description of the SVPS and its key plasma features; following this, steps for setting up the model and the flow equations are presented. It also describes the methodology for solving the equations using the boundary conditions. Section 3 presents the procedure for setting up the MCS and Section 4 gives the results and discussions. The last section is the concluding section. There are two appendices that augment the results in the paper.

## 2. Equations for a flowing magnetised plasma

### 2.1 The model

***The small volume plasma system* (SVPS):** Fig. 1(a) gives the schematic of the SVPS [44 – 47], showing the CEPS (Compact ECR Plasma Source, with its associated NdFeB ring magnets) attached to the SVPS (length ≃ 37 cm, diameter ≃ 15 cm). Fig. 1(b) shows the axial profile of the magnetic field in the expansion chamber arising due to the ring magnets of the CEPS. After the initial few cm (~ 4 cm) from the source mouth, the field falls off exponentially with a scale length, $\lambda_M \simeq 9.25$ cm. Fig. 1(c) shows the magnetic field lines and constant field contours in the expansion chamber. The field forms diverging lines outside the chamber.

Some key properties of the SVPS plasma are summarized here. (i) The electrons have constant $T_e$ and *obey the Boltzmann relation along the axis over a wide range of pressures*, implying *they are in the thermal equilibrium*. (ii) Along the axis the *normalised electron density profiles overlap very closely with the normalized magnetic field profile* exhibiting $n/B$ scaling. (iii) In addition, a separate warm electron population [45-48] was observed, that has very low density, $n_w$ ($n_w/n \sim 10^{-3}$) and fairly high temperature, $T_w$ ($T_w / T_e \sim 20 - 25$). The warm population also exhibits $n_w/B$ scaling. Its origin is believed to be linked to the plasma dynamics inside the CEPS [45].

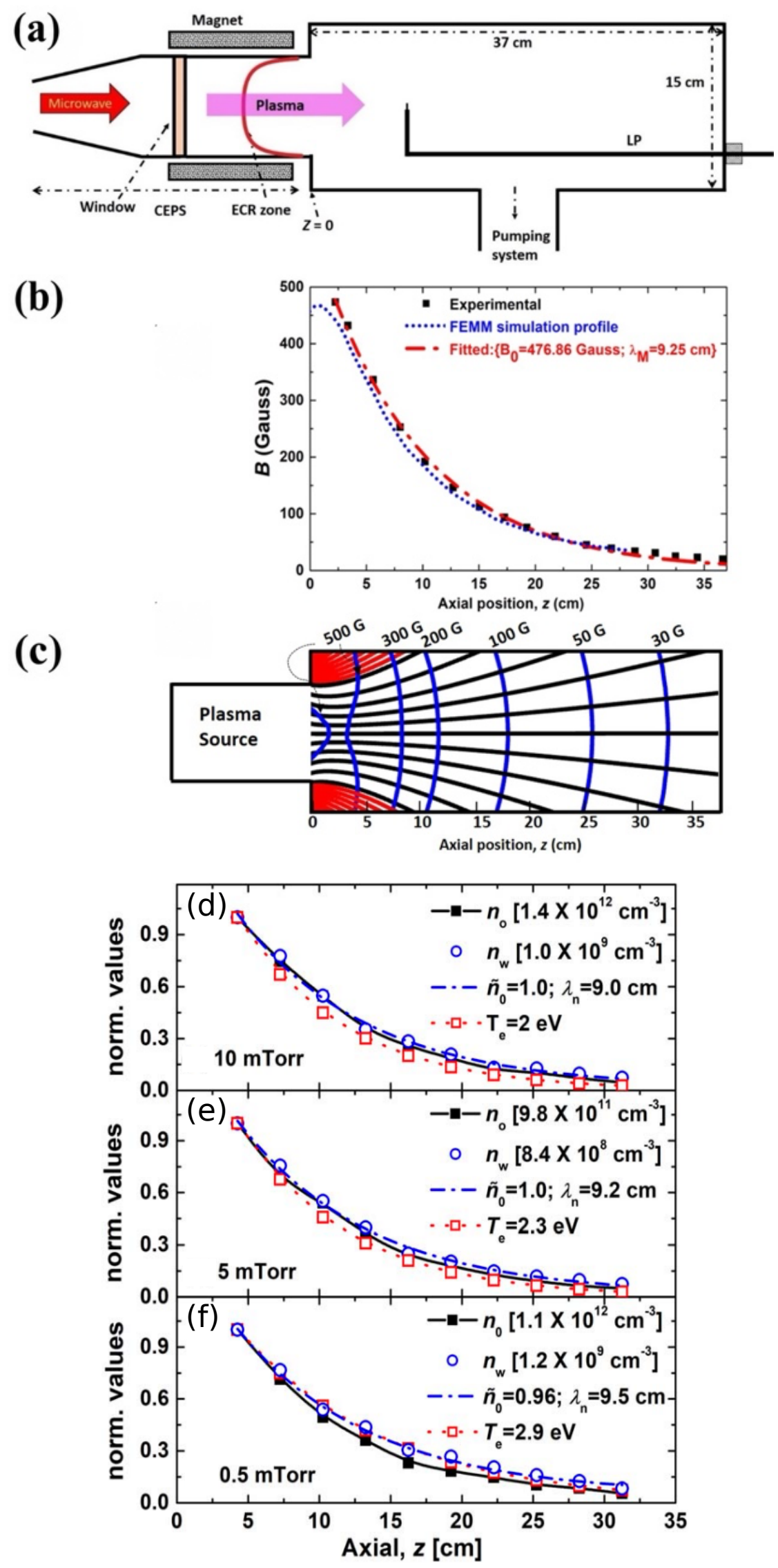

**Figure 1:** (a) Schematic of Small Volume Plasma System (SVPS); (b) Axial Magnetic field profile with exponential damping length $\lambda_M \simeq 9.25$ cm; (c) Diverging field lines in SVPS system. Figures (d), (e) and (f) give axial profiles of *normalized quantities* at ≃ 500 W power and pressures ≃10 mTorr, ≃5 mTorr and ≃0.5 mTorr, respectively. The four profiles in each subplot are as follows: (i) normalized bulk density $\tilde{n}_b$ (black solid line, solid square); (ii) normalized warm density $\tilde{n}_w$ (blue open circle) along with normalizing factor ($n_F$) given in brackets; (iii) Exponential fit of the normalized bulk density ($\tilde{n}_b$) = $n_b(z)/n_F \sim \tilde{n}_0 \exp[-(z-z_0)/\lambda_n]$ (dashed–dotted blue line) with values of $\tilde{n}_0$ and $\lambda_n$ given in legend, $z_0 = 4.25$ cm, and (iv) profile of the Boltzmann factor, $\exp(V_p(z)/T_e)$ (red dotted line, open square) with value of the constant average axial $T_e$ given in the legend.

To illustrate some of the above features, data from the SVPS experiments [48] are reproduced in Figs. 1(d) – 1(f) here. These give on-axis profiles of the (1) *normalized bulk density* $\tilde{n}_b$, (2) *normalized warm density* $\tilde{n}_w$, (3) *the Boltzmann factor* [exp $\{eV_p(z)/T_e\}$], with $V_p(z)$ being the axial plasma potential for pressures, $\simeq$ 10 mTorr, $\simeq$ 5 mTorr and $\simeq$ 0.5 mTorr. It is seen that *all three profiles practically overlap* and that *the best fits are exponentially damped profiles* [also shown in Figs. (d) – (f)] with damping lengths, $\lambda_n \simeq 9 – 9.5$ cm, over all pressures, even though the pressure varies by more than one order of magnitude. Since $T_e$ is the bulk electron temperature, it confirms the earlier claim that the bulk electrons are in thermal equilibrium. Since the axial magnetic field scale length, $\lambda_M \simeq \lambda_n$ over the entire pressure range, $\tilde{n}_b$ and $\tilde{n}_w$ *will have practically overlapping profiles with the normalized magnetic field profile B*, *suggesting robust* $n/B$ and $n_w/B$ scaling. Thus, the warm population here is confined magnetically and not by the relatively weak plasma potential in the experiments ($\simeq$ 10 – 15 V).

***Flow equations in the magnetic coordinate system* (MCS):** Consider a steady state model with cold ions ($T_i \simeq 0$). Assuming quasineutrality, i.e., $n_e \simeq n_i \simeq n$, the equations of motion for the electron and ion fluids and the equation of continuity are given below in Eq. 1. Neglecting the nonlinear inertial term, the electron momentum equation is given by Eq. (1a), where $m$, $e$, $\boldsymbol{V}$e, and $p$e, are the electron mass, charge, velocity, and the thermal pressure, respectively and $\nu_e$ is the electron-neutral elastic collision frequency. $\boldsymbol{E}$ is the self consistent, ambipolar electric field and $\boldsymbol{B}$, the applied magnetic field (due to the CEPS magnets). In the ion momentum equation (1b), $M$, $\boldsymbol{V}$, and $q$, are the ion mass, velocity and charge, respectively and $\nu_c$, the *total* (elastic and charge exchange) ion-neutral collision frequency. In the continuity equation (1c), $\nu_i$ is the frequency of ionization by electron impact on neutrals.

$$0 = -en\,[\boldsymbol{E} + \boldsymbol{V}_e \times \boldsymbol{B}] - mn\nu_e\boldsymbol{V}_e - \nabla p_e \qquad (1a)$$
$$Mn[\boldsymbol{V}\cdot\nabla\boldsymbol{V}] = qn\,[\boldsymbol{E} + \boldsymbol{V}\times\boldsymbol{B}] - Mn\nu_c\boldsymbol{V} \qquad (1b)$$
$$\nabla\cdot(n\boldsymbol{V}) = \nu_i n \qquad (1c)$$

Due to the axisymmetric nature of the CEPS and its magnetic field, $\partial/\partial\varphi \equiv 0$ ($\varphi$: azimuthal coordinate). This reduces the flow problem to a two-dimensional (2D) one in the radial and axial coordinates, $r$ and $z$, respectively, with $\varphi$ becoming irrelevant. Alternatively, one may set up a *magnetic coordinate system* (MCS) in which the $r$, $z$ coordinates are replaced by ($t$, $s$), where $t$ is the coordinate *perpendicular* to the field lines and $s$, the coordinate *along* the field lines. The azimuthal coordinate, $\varphi$ is the same in both systems. As discussed in the earlier section, the flow problem here will use the MCS coordinates ($t$, $\varphi$, $s$) instead of cylindrical coordinates ($r$, $\varphi$, $z$). Figure 2 shows the ($t$, $s$) coordinate lines or grid in the ($r$, $z$) plane. The z-direction is chosen along the system axis ($r = 0$); the system axis also corresponds to t = 0.

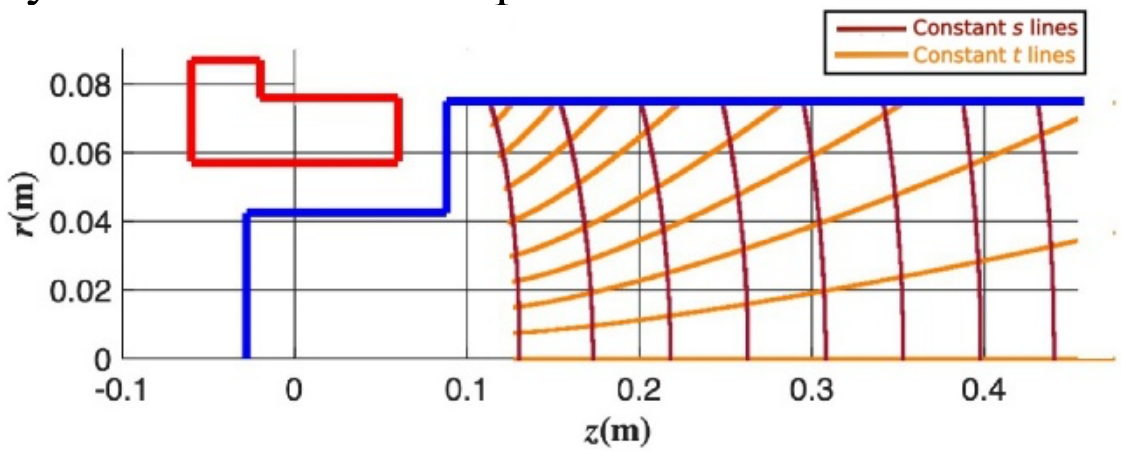

**Figure 2:** Figure shows magnetic coordinates ($t$, $s$) in the ($r$, $z$) coordinate system. The $t$ coordinate corresponds to $s$ = const. lines, and the $s$ coordinate to, $t$ = const. lines.

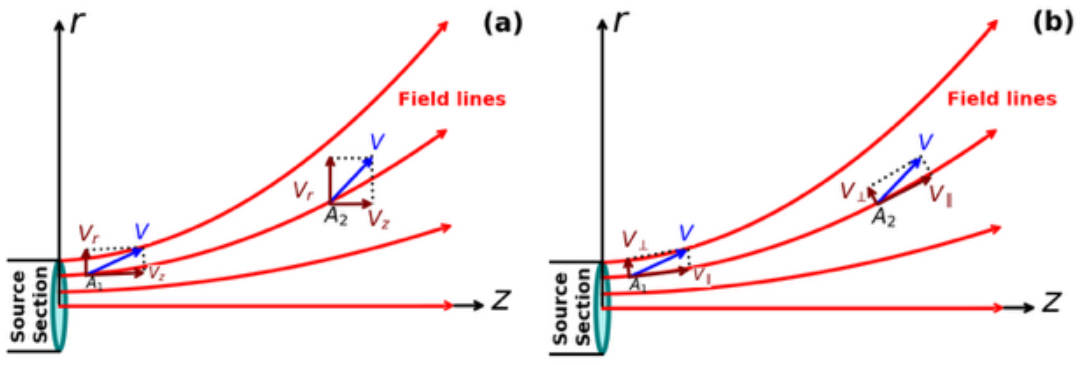

**Figure 3:** Figure shows the ordering of $\boldsymbol{V}_z$, $\boldsymbol{V}_r$ and $\boldsymbol{V}_{\parallel}$, $\boldsymbol{V}_{\perp}$ at a point of low (A$_1$) and high (A$_2$) curvature (bending) of the field lines with respect to the axis. (a) shows that the ordering of ($\boldsymbol{V}_z$, $\boldsymbol{V}_r$) changes with curvature of the field at A$_1$ and A$_2$; (b) shows that in the MCS, variables exhibit a unique ordering ($\boldsymbol{V}_{\parallel} >> \boldsymbol{V}_{\perp}$) irrespective of the curvature.

When plasma is ejected from the plasma source into the expansion chamber, it *flows primarily along the field lines aligned along the chamber axis*. Consequently, the ejected plasma continues to flow along the field lines as it proceeds into the expansion chamber. Now, *for plasma flowing mainly along a magnetic field*, *the flow velocity along the field lines is much greater than* the velocities perpendicular to the field. It turns out that this fact *offers scope for the ordering of variables* (discussed in Section 1) when the flow equations are transformed to magnetic coordinates ($t$, $\varphi$, $s$). To see why this is relevant, one may compare the *ordering of the flow velocity variables*, $V_z$ and $V_r$ (in cylindrical coordinates) at *two* points on the field line, one being a point of *low curvature with weak bending* (A$_1$) and the other (A$_2$), being a point of *high curvature with strong bending*. Figure 3(a) shows that at A$_1$ (field ~ along $z$ direction), $|V_z| > |V_r|$, whereas at A$_2$ (field ~ bending towards radial direction) the reverse is true, i.e., $|V_r| > |V_z|$. In fact, the inequality will flip each time the field line changes direction from axial to radial direction or vice versa. In contrast, Figure 3(b) presents the scenario in the MCS velocity variables, $V_{\parallel}$ (along the field line) and $V_{\perp}$ (perpendicular to the field line), where the inequality, $|V_{\parallel}| >> |V_{\perp}|$ holds not only at $A_1$ and $A_2$, but at all points of the field line irrespective of how the field line bends or curves. *It is apparent therefore, that the MCS is a natural coordinate system in which to formulate the flow problem*: *As noted in Section 1, the MCS allows the flow variables to be partitioned into a lowest order and a higher order group. Furthermore, using these ordered variables, one may split each equation of the model into a pair of reduced or simpler equations, that are easier to solve with the available boundary conditions.*

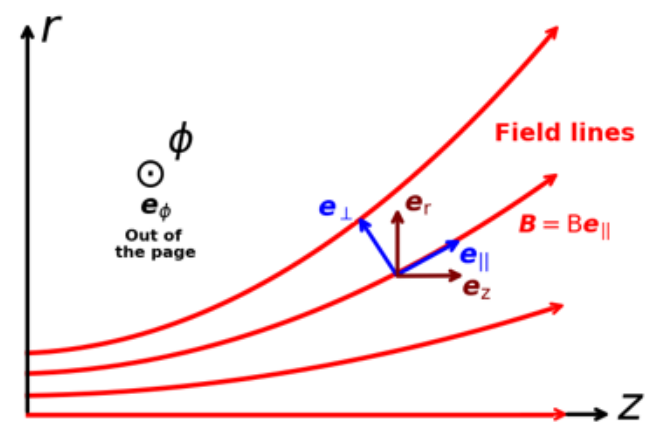

**Figure 4:** The unit vectors *along* the field and *perpendicular* to the field have been shown as $\boldsymbol{e}_{\parallel} = b_r \boldsymbol{e}_r + b_z \boldsymbol{e}_z$ and $e_{\perp} = b_z \boldsymbol{e}_r - b_r \boldsymbol{e}_z$. The magnetic field is expressed as $\boldsymbol{B} = B\,\boldsymbol{e}_{\parallel} = B_r\,\boldsymbol{e}_r + B_z\,\boldsymbol{e}_z$. Here, $b_r = B_r/B$, $b_z = B_z/B$ and $B = [B_r^2 + B_z^2]^{1/2}$. $B$, $B_r$ and $B_z$ represent the total, $r$ and $z$ components of the magnetic field.

***Flow equations in magnetic coordinates:*** Figure 4 shows the basis vectors $(\boldsymbol{e}_{\perp}, \boldsymbol{e}_{\parallel})$ in the MCS with $\boldsymbol{e}_{\perp}$ being perpendicular to $\boldsymbol{B}$ and $\boldsymbol{e}_{\parallel}$, parallel to $\boldsymbol{B}$. Thus, the MCS uses the triplet of unit vectors $(\boldsymbol{e}_{\perp}, \boldsymbol{e}_{\varphi}, \boldsymbol{e}_{\parallel})$ in the $(t, \varphi, s)$ coordinates, with $(\boldsymbol{e}_{\perp}, \boldsymbol{e}_{\parallel})$ being defined as,

$$\boldsymbol{e}_{\perp} = b_z\,\boldsymbol{e}_r - b_r\,\boldsymbol{e}_z;\ \boldsymbol{e}_{\parallel} = b_r\,\boldsymbol{e}_r + b_z\,\boldsymbol{e}_z \qquad (2)$$

where, $b_r = B_r/B$, $b_z = B_z/B$; $\boldsymbol{B} = B_r\boldsymbol{e}_r + B_z\boldsymbol{e}_z = B\,\boldsymbol{e}_{\parallel}$; $B = [B_r^2 + B_z^2]^{1/2}$. $B_r$ and $B_z$ are the $r$- and and $z$-components of the magnetic field. The transformation of the flow equations from the $(r, z)$ to the $(t, s)$ coordinates including formulae for the transformation of derivatives, etc., are derived in Appendix A.

***Flow equations for electrons:*** There is no *energy transport equation for electrons* in Eqns. (1). Thus, an equation for the *thermodynamic path* for electrons is needed. As discussed earlier, $T_e \simeq$ *const.* along the axis, with the bulk electrons obeying the Boltzmann relation, *exhibiting* $n/B$ *scaling* (see also Sec. 4.2) *and obeying the double adiabatic equation of state* (see [48]). The latter two are sensitive and rarely observed properties and may be assumed to hold in a limited region around the axis. On the other hand, the isothermal property of electrons can be expected to hold more robustly. Thus, *in the present work, the bulk electrons are assumed isothermal in the entire plasma*, which would also provide the required thermodynamic path.

It should also be mentioned that *due to its very low density* $n_w$, *the warm population is not expected to participate in the ion flow dynamics of the system*. It can however, due to its high $T_w$, ionize the background gas, which would mainly affect the *cross-field plasma density profiles*; the profiles along the field lines are expected to remain largely unaffected since these obey either $n/B$ scaling (see Sec. 4.2) or follow the ion density profiles. It will be seen later that the measured radial profiles do show slight shift from the profiles calculated from the model. To conclude, in the flow dynamics model being developed below, the warm population will not be included.

In a steady state, the ambipolar electric field $\boldsymbol{E}$ obeys, $\nabla \times \boldsymbol{E} \simeq 0$ so that $\boldsymbol{E}_{\varphi} \simeq 0$. Dropping the collision or friction term from the three equations of (1a) gives on using $\boldsymbol{B} = B\,\boldsymbol{e}_{\parallel}$,

⊥ Comp:

$$0 \simeq -enE_{\perp} - enV_{e\varphi}B - \frac{\partial p_e}{\partial t}$$
$$\Rightarrow V_{e\varphi} \simeq -\frac{E_{\perp}}{B} - \frac{1}{enB}\frac{\partial p_e}{\partial t} \qquad (3)$$

$\varphi$ Comp:

$$enV_{e\perp}B \simeq 0 \Rightarrow V_{e\perp} \simeq 0 \qquad (4)$$

|| Comp:

$$0 \simeq -enE_{\parallel} - \frac{\partial p_e}{\partial s} \qquad (5)$$

In the second equation of (3) the first term is the $\boldsymbol{E} \times \boldsymbol{B}$ drift and the second, the diamagnetic drift; both these are along $\boldsymbol{e}_{\varphi}$. Note that no such term arises along $\boldsymbol{e}_{\parallel}$. Because $\nabla \times \boldsymbol{E} \simeq 0$, one may use $\boldsymbol{E} = -\nabla\Phi$ ($\Phi$: electrostatic potential), in Eqs. (3) and (5). However, this will not lead to any simplification unless $T_e$ is a constant (thermal equilibrium). Recalling, however, that the electrons in the SVPS have been taken to be isothermal, one may use $\boldsymbol{E} = -\nabla\Phi$ and $p_e \simeq neT_e$ in (5) to obtain,

$$E_{\parallel} = -\partial\Phi/\partial s = -T_e\frac{\partial \ln n}{\partial s} \qquad (6)$$

*s – integration* of (6) yields,

$$\Phi = T_e \ln n + f(t) + C_1 \qquad (7)$$

where $f$ is a function of $t$ alone and $C_1$ is a constant. One may use (7) to determine $E_\perp = -\,\partial\Phi/\partial t$ and substitute for $E_\perp$ in (3) to obtain,

$$\frac{df}{dt} \simeq V_{e\varphi}B \tag{8}$$

The LHS of (8) is a function of $t$ alone, whereas the RHS is in general, a function of both $t$ and $s$, implying that $\frac{df}{dt} \simeq$ *constant*. It suffices to note that because $V_{e\varphi} \simeq 0$ on the chamber axis ($t$ = 0), the constant is zero. Thus,

$$\frac{df}{dt} \simeq 0$$
$$\Rightarrow V_{e\varphi} \simeq 0 \textit{ everywhere, for all } t \textit{ and } s. \tag{9}$$

$\frac{df}{dt} \simeq 0$ implies $f(t)$ is a constant. The latter can be absorbed in the constant $C_1$ in (7) without loss of generality. Using (9) in (3) and (5) obtains the following equations for $E_\perp$ and $E_{||}$,

$$E_\perp = -\,T_e\frac{\partial \ln n}{\partial t}, \quad E_{||} = -\,T_e\frac{\partial \ln n}{\partial s} \tag{10}$$

***Remarks:*** The electron diamagnetic current, $\boldsymbol{J}_e = J_{e\varphi}\,\boldsymbol{e}_\varphi$ gives a force, $\boldsymbol{J}_e \times \boldsymbol{B} = J_{e\varphi}B\,\boldsymbol{e}_\perp = -\,enV_{e\varphi}B\,\boldsymbol{e}_\perp$, that has components along the z-direction as well as the radial direction. In general, the component along the $z$-direction adds to the thrust by the ions. However, in the present case, the isothermal electrons of the SVPS, render $J_{e\varphi} = -enV_{e\varphi} \simeq 0$ and a vanishing axial thrust due to the diamagnetic electron current.

***Flow equations for ions:*** The components of the flow equations for ions are obtained from Eq. (1b). Opening out the expression for the inertial term $[\boldsymbol{V}\cdot\boldsymbol{\nabla}\boldsymbol{V}]$ in in ($t$, $s$) coordinates (see Appendix A), obtains the following flow equations for *singly* charged ions.

**⊥ - Component:**

$$V_\perp\frac{\partial V_\perp}{\partial t} - a_1 V_{||}V_\perp - b_z\frac{V_\phi^2}{r} + V_{||}\frac{\partial V_\perp}{\partial s} - a_2 V_{||}^2 + \nu_c V_\perp = \frac{e}{M}E_\perp + \omega_{ci}V_\phi - \nu_i V_\perp \tag{11}$$

***ϕ* - Component:**

$$V_\perp\frac{\partial V_\phi}{\partial t} + b_z\frac{V_\phi V_\perp}{r} + b_r\frac{V_\phi V_{||}}{r} + V_{||}\frac{\partial V_\phi}{\partial s} + \nu_c V_\phi = -\omega_{ci}V_\perp - \nu_i V_\phi \tag{12}$$

**|| - Component:**

$$a_1 V_\perp^2 + V_\perp\frac{\partial V_{||}}{\partial t} - b_r\frac{V_\phi^2}{r} + a_2 V_\perp V_{||} + V_{||}\frac{\partial V_{||}}{\partial s} + \nu_c V_{||} = \frac{e}{M}E_{||} - \nu_i V_{||} \tag{13}$$

In Eqs. (11) – (13), $V_\varphi$ is the ion flow velocity in the azimuthal direction; $a_1$ and $a_2$ are *measures of the normalized magnetic field scale length along t and s, respectively* and are defined in Eqs. (A-11), (A-14), (A-15) and $\omega_{ci} = eB/M$ is the ion cyclotron frequency. In equations (11) – (13) the last term gives the drag force acting on the ion fluid as the newly born ions are accelerated to the ion fluid velocity, $\boldsymbol{V}$. The continuity equation (1c) in ($t$, $s$) coordinates reads:

**Continuity equation:**

$$\frac{\partial V_\perp}{\partial t} + V_\perp\frac{\partial \ln n}{\partial t} + \frac{\partial V_{||}}{\partial s} + V_{||}\frac{\partial \ln n}{\partial s} + \left[-a_1 + \frac{b_r}{r}\right]V_{||} + \left[\frac{b_z}{r} + a_2\right]V_\perp = \nu_i \tag{14}$$

***Normalization:*** It is convenient to work with normalized variables. The normalizations for the different quantities are listed in Table 1. The plasma density is normalized with respect to the highest density, $n_{max}$ in the experiments. The ion velocities are normalized with respect to a *normalization Bohm speed* $V_{B0} = [eT_{e0}/M]^{1/2}$, where $T_{e0}$ is the electron temperature at a suitable point in the system. In the SVPS experiments under consideration here, $T_e$ is a constant and so, $T_{e0} = T_e$. The normalized ion velocities are designated by $U$. The *normalized* Bohm velocity, $U_B = V_B / V_{B0} = 1$, since $T_e$ is a constant.

The ion – neutral collision frequency, $\nu_c = N_g\sigma_c V_{||}$. Here, $V_{||}$ is used since it is the highest speed ions acquire (for cold ions) and hence, also the speed with which they collide (variable mobility scenario [22]). $\sigma_c$ is the total differential cross section for ion – neutral collisions (including elastic and charge exchange) and $N_g$, the gas density. The mean free path $\lambda_c = [N_g\sigma_c]^{-1}$, which gives $\nu_c = \frac{V_{||}}{\lambda_c}$ and so after division by $V_{B0}$ (normalization), $\nu_c \rightarrow \frac{\nu_c}{V_{B0}} = \frac{U_{||}}{\lambda_c}$. The ionization frequency, $\nu_i = N_g R_{ion}$, where $R_{ion}$ is the rate constant for ionization by electron impact. Upon normalization: $\nu_i \rightarrow \frac{\nu_i}{V_{B0}} = \frac{1}{L_i}$; also, $\omega_{ci} \rightarrow \frac{\omega_{ci}}{V_{B0}} = \frac{1}{R_{LB}}$. Applying normalization using Table 1 yields the normalized equations.

**Table 1: Normalization of plasma variables.**

| No. | Variables | Normalization | Normalized Variables |
|---|---|---|---|
| 1 | Density | Maximum Density ($n_{max}$) | $n$ |
| 2 | Bohm Velocity ($V_B$) | Bohm Velocity: $V_{B0} = \sqrt{\frac{eT_{e0}}{M}}$; $T_{e0}$: Suitable Constant Temperature | $U_B$ (= 1) [In SVPS expts.] |
| 3 | Ion Velocity $(V_{||}, V_\perp, V_\phi)$ | Maximum Bohm Velocity ($V_{B0}$) | $(U_{||}, U_\perp, U_\phi)$ |

**⊥ Component:**

$$U_\perp\left[\frac{\partial U_\perp}{\partial t} - a_1 U_{||}\right] + U_{||}\left[\frac{\partial U_\perp}{\partial s} - a_2 U_{||}\right] + \frac{U_{||}U_\perp}{\lambda_c} - U_\phi\left[\frac{1}{R_{LB}} + \frac{b_z U_\phi}{r}\right] = -U_B^2\frac{\partial \ln n}{\partial t} - \frac{U_\perp}{L_i} \quad (15)$$

**|| Component:**

$$U_\perp\left[\frac{\partial U_{||}}{\partial t} + a_1 U_\perp\right] + U_{||}\left[\frac{\partial U_{||}}{\partial s} + a_2 U_\perp\right] + \frac{U_{||}^2}{\lambda_c} - \frac{b_r U_\phi^2}{r} = -U_B^2\frac{\partial \ln n}{\partial s} - \frac{U_{||}}{L_i} \quad (16)$$

***ϕ* Component:**

$$U_\perp\left[\frac{\partial U_\phi}{\partial t} + \frac{b_z}{r}U_\phi + \frac{1}{R_{LB}}\right] + \frac{U_\phi U_{||}}{\lambda_c} + U_{||}\left[\frac{\partial U_\phi}{\partial s} + \frac{b_r}{r}U_\phi\right] + \frac{U_\phi}{L_i} = 0 \quad (17)$$

**Continuity equation:**

$$\frac{\partial U_\perp}{\partial t} + U_\perp\frac{\partial \ln n}{\partial t} + \frac{\partial U_{||}}{\partial s} + U_{||}\frac{\partial \ln n}{\partial s} + \left[-a_1 + \frac{b_r}{r}\right]U_{||} + \left[\frac{b_z}{r} + a_2\right]U_\perp = \frac{1}{L_i} \quad (18)$$

***Need for splitting equations:*** Eqns. (15) – (18) involve partial derivatives of all variables ($U_\perp$, $U_\phi$, $U_{||}$, $n$) with respect to $t$ and $s$. *To solve these equations over a relevant portion of the t - s plane one must integrate along chosen paths of integration that may be parametrised using a suitable parameter to translate the paths for scanning the plane*. For instance, if one integrates along different *t*-coordinate lines, the associated values of *s* that label the paths may be used as the parameter to move the paths over the plane. Likewise, to integrate along different *s*-coordinate lines, the associated values of *t* may be used to shift the paths over the plane.

Each integration along a path must be initiated from a boundary point. Thus, for an "ensemble of such paths" the associated set of boundary points would yield a corresponding boundary line. *Consider the situation that the integrations are undertaken along different s-coordinate lines*. In this case, *the boundary line will correspond to a suitable t-coordinate line* (*or s = constant contour*) along which all the variables and their partial derivatives with respect to *t* have to be specified (the partial derivatives with respect to *s* being determined by the model equations themselves). Now, the boundary conditions are derived from experimental measurements, which are undertaken *usually along the system axis and along the radius at a few axial locations*. Since neither the axis nor the radial direction coincides with any *t-coordinate or s = constant line*, it would not be possible to specify the variables and their derivatives along such a boundary. Moreover, the data is acquired only at a few discrete points, which would render evaluation of the derivatives highly inaccurate.

On the other hand, if integration proceeds along the *t-coordinate lines*, integration maybe initiated from $t = 0$ (the axis) as the boundary line and all variables along with their partial derivatives with respect to *s* have to be specified along this line (the partial derivatives with respect to *t* being determined by the equations of the model). As mentioned above, on-axis measurement of variables is most common since it is simpler and fairly accurate. The measured variables are typically, the plasma density *n*, the electron temperature $T_e$, the plasma potential *Φ* and the parallel ion velocity $U_{||}$, although in the present work the latter was not available. In addition, *kinematic restrictions fix the on-axis values of* $U_\phi$ *and* $U_\perp$ *to zero*.

Integration along a *t*-coordinate also requires the partial derivatives, $\frac{\partial U_{||}}{\partial s}$, $\frac{\partial \ln n}{\partial s}$, $\frac{\partial U_\perp}{\partial s}$, $\frac{\partial U_\phi}{\partial s}$ to be specified along the axis ($t$ = 0). The last two are identically zero by virtue of both $U_\phi$ and $U_\perp$ being zero on the axis. However, the first two ($\frac{\partial U_{||}}{\partial s}$, $\frac{\partial \ln n}{\partial s}$) remain undetermined since $U_{||}$ and $n$ are measured only at few points on the axis, from which it is not possible to obtain unique and accurate profiles of the derivatives. This is a serious issue. Moreover, *to start integrations along t-coordinates from the axis* ($t$ = 0) one encounters singularities on the axis. *It is here that splitting of the model equations helps, since both issues can be solved using the method described below.*

***Ordering of variables:*** It is worth noting that aside from the primary or basic ordering, $|U_{||}| >> |U_\perp| \sim |U_\phi|$, *no a priori ordering sequence is available for the other variables of the flow problem even in the magnetic coordinate system*. *Nonetheless, based on this basic ordering and the likely nature of the other variables*, it is possible to build further on the primary ordering to partition the remaining variables and their derivatives into one of the two groups – a lowest order group, or a higher order group. As an *initial guess*, the following *a priori* ordering was used.

***Lowest Order Variables* (LO):** $U_{||}$, ln ($n$), $\frac{\partial U_{||}}{\partial s}$, $\frac{\partial \ln n}{\partial s}$, $\frac{\partial \ln B}{\partial s}$, $a_1$, etc., … (19)

***Higher Order Variables* (HO):** $U_\perp$, $U_\phi$, $\frac{\partial U_\perp}{\partial s}$, $\frac{\partial U_\phi}{\partial s}$, $\frac{\partial U_\perp}{\partial t}$, $\frac{\partial U_\phi}{\partial t}$, $\frac{\partial U_{||}}{\partial t}$, $\frac{\partial \ln B}{\partial t}$, $a_2$, etc.,… (20)

***Remark:*** Thus, apart from $U_{||}$ one also treats ln ($n$) and the derivatives of $U_{||}$, ln ($n$), ln ($B$), etc., along *s* as LO variables. This classification or grouping is based on the premise that similar to $U_{||}$, ln($n$), etc., derivatives of such terms along *s* would also tend to dominate. One has $-a_1 + \frac{b_r}{r} = \frac{\partial \ln B}{\partial s}$, so that one may classify $a_1$ as an LO variable since $\frac{\partial \ln B}{\partial s}$ has been classified as one. *All remaining variables are grouped as* HO. Thus, in addition to the *basic* HO variables, $U_\perp$, $U_\phi$, $\frac{\partial U_\perp}{\partial s}$, $\frac{\partial U_\phi}{\partial s}$, $\frac{\partial U_\perp}{\partial t}$, $\frac{\partial U_\phi}{\partial t}$, *derivatives of* LO *variables with respect to t* are also to be placed in the HO category, since variations perpendicular to the field lines are expected to be small. Thus, $a_2$ falls in the HO category as well because $a_2 = -\frac{\partial \ln B}{\partial t}$.

***Ordering of the product terms in the equations:*** It may be borne in mind at the outset that *no systematic scheme that assigns numerical values to the orders of the variables, is available*. Nonetheless, an ordering is required for splitting equations in a manner that permits solution of the resulting equations relatively easily. A little consideration shows that this would be achieved most simply if the partial derivatives of $U_\parallel$, $U_\perp$ and $U_\phi$ with respect to *t* ($\frac{\partial U_\perp}{\partial t}$, $\frac{\partial U_\parallel}{\partial t}$, $\frac{\partial U_\phi}{\partial t}$) and *s* ($\frac{\partial U_\perp}{\partial s}$, $\frac{\partial U_\parallel}{\partial s}$, $\frac{\partial U_\phi}{\partial s}$) could be separated into *different* equations. Thus, one could attempt to split the equations so that *only two equations result from each*. One notes however, that the equations also contain terms involving products of variables. It is therefore, *important that the ordering of the product terms be undertaken in a manner that the primary objective of splitting each equation into two equations is not thwarted.* Keeping these requirements in mind, the *following ansatz* was adopted to arrive at a *tentative ordering scheme for the product terms*.

- **(i)** LO × LO ~ $LO^2$
- **(ii)** LO × HO ~ $LO^{0.5}$ ~ IO, asserting that HO ~ $LO^{-0.5}$
- **(iii)** LO × IO ~ $LO^{1.5}$
- **(iv)** LO × $HO^2$ ~ $LO^{0.5}$ × HO ~ $LO^0$ ~ $HO^0$ and so on.
- **(v)** IO: *Intermediate order*
- **(vi)** Step (ii) above, is a crucial step in the ordering scheme for product terms. *Since the orders of the variables do not have any preassigned numerical values, this step sets up the connection between the* HO *and the* LO *variables so that product terms such as* LO × HO, *etc., arising in the equations get grouped as per expected behavior of the variables.*
- **(vii)** Application of the above ansatz *splits each equation into two groups*. One group, labeled as LO, contains only LO terms: All terms in this group have *positive* powers of LO and are treated on par simply as LO, irrespective of the power. A similar picture holds for the HO set.
- **(viii)** Step (iv) yields terms like $LO^0$ ~ $HO^0$ that cannot be classified in either category, *a priori*. *However, their structure and trial / testing can help classify such terms as either LO or HO, but not both.*

***Remark:*** *It must be stressed here the above scheme is tentative and its validity has to be justified by the results derived from the model ultimately.*

**⊥ Component:**

$$U_\parallel\left[\frac{\partial U_\perp}{\partial s} - a_1 U_\perp - a_2 U_\parallel + \frac{U_\perp}{\lambda_c}\right] + \left[U_\perp \frac{\partial U_\perp}{\partial t} + \frac{U_\perp}{L_i} + \frac{\partial \ln n}{\partial t}\right] - U_\phi\left[\frac{1}{R_{LB}} + \frac{b_z U_\phi}{r}\right] = 0 \quad (21)$$

$LO^{0.5} + LO^{1.5} + LO^{1.5} + LO^{0.5} + HO^2 + HO + HO + HO + HO^2$

**|| Component:**

$$U_\parallel\left[\frac{\partial U_\parallel}{\partial s} + \frac{U_\parallel}{\lambda_c} + \frac{1}{L_i} + \frac{\partial \ln n}{\partial s}\right] + U_\perp\left[\frac{\partial U_\parallel}{\partial t} + a_2 U_\parallel + a_1 U_\perp\right] - \frac{b_r U_\phi^2}{r} = 0 \quad (22)$$

$LO^2 + LO^2 + LO + LO + HO^2 + HO^0 + HO^{1.5} + HO^2$

***ϕ* Component:**

$$U_\parallel\left[\frac{\partial U_\phi}{\partial s} + \frac{b_r}{r} U_\phi\right] + \frac{U_\phi U_\parallel}{\lambda_c} + U_\perp\left[\frac{\partial U_\phi}{\partial t} + \frac{b_z}{r} U_\phi + \frac{1}{R_{LB}}\right] + \frac{U_\phi}{L_i} = 0 \quad (23)$$

$LO^{0.5} + LO^{0.5} + LO^{0.5} + HO^2 + HO^2 + HO + HO$

**Continuity equation:**

$$\left[\frac{\partial U_\parallel}{\partial s} + U_\parallel \frac{\partial \ln n}{\partial s} + \left(-a_1 + \frac{b_r}{r}\right) U_\parallel\right] + \left[\frac{\partial U_\perp}{\partial t} + U_\perp \frac{\partial \ln n}{\partial t} + \left(a_2 + \frac{b_z}{r}\right) U_\perp - \frac{1}{L_i}\right] = 0 \quad (24)$$

$LO + LO^2 + LO^2 + HO + HO^2 + HO^2 + HO + HO$

In (24) above, $-a_1 + \frac{b_r}{r} = \frac{\partial \ln B}{\partial s}$ and so, the third LO term in the 1st bracket qualifies as LO. The inhomogeneous term, $\frac{1}{L_i}$ is n for evaluating the limiting value of $\frac{b_z}{r} U_\perp$ as, $U_\perp \to 0$ and $r(t) \to 0$ as $t \to 0$. In (22), the term $HO^0$ has been classified as HO, as this turns out to be the correct choice in this case [see note (viii) above]. Splitting the LO and HO terms into separate equations yield:

**⊥ Component:**

$$\frac{\partial U_\perp}{\partial s} - a_1 U_\perp - a_2 U_\parallel + \frac{U_\perp}{\lambda_c} \simeq 0 \quad (25)$$

$$U_\perp \frac{\partial U_\perp}{\partial t} + \frac{U_\perp}{L_i} + \frac{\partial \ln n}{\partial t} - \frac{U_\phi}{R_{LB}} - \frac{b_z U_\phi^2}{r} \simeq 0 \quad (26)$$

**|| Component:**

$$U_\parallel \frac{\partial U_\parallel}{\partial s} + \frac{U_\parallel^2}{\lambda_c} + \frac{U_\parallel}{L_i} + \frac{\partial \ln n}{\partial s} \simeq 0 \quad (27)$$

$$U_{\perp}\frac{\partial U_{||}}{\partial t} + a_2 U_{||}U_{\perp} + a_1 U_{\perp}^2 - \frac{b_r U_{\phi}^2}{r} \simeq 0 \qquad (28)$$

**$\phi$ Component:**

$$U_{||}\left[\frac{\partial U_{\phi}}{\partial s} + \frac{b_r}{r}U_{\phi}\right] + \frac{U_{\phi}U_{||}}{\lambda_c} \simeq 0 \qquad (29)$$

$$U_{\perp}\left[\frac{\partial U_{\phi}}{\partial t} + \frac{b_z}{r}U_{\phi} + \frac{1}{R_{LB}}\right] + \frac{U_{\phi}}{L_i} \simeq 0 \qquad (30)$$

**Continuity equation:**

$$\frac{\partial U_{||}}{\partial s} + U_{||}\frac{\partial \ln n}{\partial s} + \left(-a_1 + \frac{b_r}{r}\right)U_{||} \simeq 0 \qquad (31)$$

$$\frac{\partial U_{\perp}}{\partial t} + U_{\perp}\frac{\partial \ln n}{\partial t} + \left(a_2 + \frac{b_z}{r}\right)U_{\perp} - \frac{1}{L_i} \simeq 0 \qquad (32)$$

***Remark:*** The $\simeq$ sign in Eqns. (25) – (32) emphasises the approximate nature of the equations. These equations, along with Eq. (10) comprise the complete set of equations for the flow model. These will have to be solved for the SVPS using data from the experiments. Two major advantages have resulted. (i) By considering selected sets of equations one may bypass the requirement of having to evaluate partial derivatives with respect to *s*. (ii) The singularities are also eliminated and one needs to evaluate only a few indeterminate (zero / zero) terms in the limit, $t \to 0$.

It is convenient to decouple $\frac{\partial U_{\perp}}{\partial t}$ and $\frac{\partial \ln n}{\partial t}$ in Eqns. (26) and (32). Doing this yields the equations,

$$\frac{\partial U_{\perp}}{\partial t} + \frac{1}{1-U_{\perp}^2}\left[\left(\frac{b_z}{r} + a_2\right)U_{\perp} - \frac{1}{L_i}(1+U_{\perp}^2) + \frac{U_{\perp}U_{\phi}}{R_{LB}} + \frac{b_z U_{\perp}U_{\phi}^2}{r}\right] \simeq 0 \qquad (33)$$

$$\frac{\partial \ln n}{\partial t} + \frac{U_{\perp}}{1-U_{\perp}^2}\left[\left(\frac{b_z}{r} + a_2\right)U_{\perp} - \frac{1}{L_i}(1+U_{\perp}^2) + \frac{U_{\perp}U_{\phi}}{R_{LB}} + \frac{b_z U_{\perp}U_{\phi}^2}{r}\right] + \frac{U_{\perp}}{L_i} - \frac{U_{\phi}}{R_{LB}} - \frac{b_z U_{\phi}^2}{r} \simeq 0 \qquad (34)$$

### 2.2 Analysis and Solution of the Model Equations

***Boundary conditions and indeterminate terms on the axis (t = 0):*** $n(0,s)$, $T_e(0,s)$ and $\Phi(0,s)$ data obtained from experiments give the boundary conditions for these variables along the $t = 0$ axis. It was noted earlier that $T_e$ would be assumed a constant throughout the plasma and that it was not possible to measure $U_{||}(t=0,s)$. It will be seen below however, that for the SVPS, $U_{||}(t=0,s)$ is *constant on the axis*. The rationale for this and the determination of $U_{||}$ will be discussed below. Collecting the boundary conditions gives,

**(1)** $U_{\perp}(0,s) = U_{\phi}(0,s) = 0$

$$\Rightarrow \quad \frac{\partial U_{\perp}}{\partial s}(0,s) = \frac{\partial U_{\phi}}{\partial s}(0,s) = 0 \qquad (35)$$

**(2)** $U_{||}(0,s) = constant$

$$\Rightarrow \quad \frac{\partial U_{||}}{\partial s}(0,\ s) = 0 \qquad (36)$$

**(3)** $T_e$ = *constant everywhere*

For integrations along the *t*-coordinate the relevant equations are, (28), (30), (33), and (34). These equations have indeterminate (0/0) terms, whose limiting values need to be determined.

**(4)** In (33) one sees that the terms, $\frac{b_z U_{\perp}}{r}$ and $b_z\frac{U_{\perp}U_{\phi}^2}{r}$ are indeterminate at $t = 0$. One may determine their limiting values treating *r* as function of *t* and using formulas given in Appendix A. Thus, (33) yields (dropping the 2nd argument),

$$t \to 0,\ \frac{\partial U_{\perp}}{\partial t}(t) = \frac{\partial U_{\perp}}{\partial t}(0) = \frac{1}{2L_i} \qquad (37)$$

**(5)** In (30) the terms, $\frac{b_z}{r}U_{\phi}$ and $\frac{U_{\phi}}{L_i\,U_{\perp}}$ are indeterminate at $t = 0$. Evaluating their limiting values (30) gives

$$t \to 0,\ \frac{\partial U_{\phi}}{\partial t}(t) = \frac{\partial U_{\phi}}{\partial t}(0) = -\frac{1}{4\,R_{LB}} \qquad \text{`}(38)$$

**(6)** Similarly (28), after evaluation of the term $\frac{U_{\phi}^2\, b_r}{U_{\perp}r}$, gives

$$t \to 0,\ \frac{\partial U_{||}}{\partial t}(t) = \frac{\partial U_{||}}{\partial t}(0) = 0 \qquad (39)$$

Finally, in (34) all the indeterminate terms $b_z\frac{U_{\perp}^2}{r}$, $\frac{b_z U_{\perp}^2 U_{\phi}^2}{r}$ and $\frac{b_z U_{\phi}^2}{r}$ vanish in the limit $t \to 0$, so that (34) gives,

$$t \to 0,\ \frac{\partial \ln n}{\partial t}(t) = \frac{\partial \ln n}{\partial t}(0) = 0 \qquad (40)$$

***Conservation of $\left(\frac{nU_{||}}{B}\right)$ along a field line:*** It was seen earlier that, $-a_1 + \frac{b_r}{r} = \frac{\partial \ln B}{\partial s}$. Substituting into (31) gives on simplifying,

$$\frac{\partial \ln\left(\frac{nU_{||}}{B}\right)}{\partial s} \simeq 0 \qquad (41)$$

Using elementary arguments, it can be shown that the relation $\left(\frac{nU_{||}}{B}\right) \simeq$ *const.*, *holds within a flux tube*, which expresses the fact that the *ion particle flux is conserved within a flux tube*, as is the magnetic flux [22]. Eq. (41) may be written as,

$$\ln\left(\frac{nU_{||}}{B}\right) \simeq C_1 \quad \Rightarrow \quad \frac{nU_{||}}{B} \simeq C \qquad (42)$$

*Thus, Eq. (42) asserts the approximate constancy of $\frac{nU_{||}}{B}$ along the field lines. This is a stronger requirement.*

***Evaluation of $U_{||}$ on the axis*:** It was noted earlier that the plasma density for the SVPS obeys $\frac{n}{B}$ scaling or the relation, $\frac{n}{B} \simeq$ const. holds along the axis (see also Fig 8, Sec. 4.1). *The approximate constancy of* $\frac{n}{B}$ *on the* system *axis (derived from the density and magnetic field data) therefore, serves as an added boundary condition.* One may write (42) as,

$$\ln\left(\frac{nU_{||}}{B}\right) = \ln\frac{n}{B} + \ln U_{||} \simeq C_1$$
$$\Rightarrow \quad \ln U_{||} \simeq const$$
$$\Rightarrow \quad U_{||} \simeq const \qquad (43)$$

The second step follows from the constancy of $\frac{n}{B}$ or $\ln\frac{n}{B}$ *on the axis*, so that $U_{||}$ is also approximately constant there. As shown in Fig. 1(b), the on-axis magnetic field obeys,

$$B \sim \exp\left(-\frac{s}{\lambda_M}\right) \quad \text{or} \quad n \sim \exp\left(-\frac{s}{\lambda_M}\right) \qquad (44)$$

Where $\lambda_M \sim 9.25$ cm. In the second step of (44), $\frac{n}{B}$ scaling has been invoked and *z* has been replaced by *s* (valid on the axis). Noting that (43) demands, $\frac{\partial U_{||}}{\partial s}(0,s) \simeq 0$, one may invoke (44) to write Eq. (27) as,

$$\frac{U_{||}^2}{\lambda_c} + \frac{U_{||}}{L_i} - \frac{1}{\lambda_M} \simeq 0 \quad (t = 0) \qquad (45)$$

Solving (45) gives $U_{||}$ on the axis.

***Evaluation of $U_{||}$ away from the axis ($t \neq 0$):*** After evaluation of $U_\perp$, *n* and $U_\phi$ (see below) one may integrate (28) to determine $U_{||}$ along the *t* coordinate lines for *different s* = constant values in the MCS coordinate grid, starting from *t* = 0, using the value of $U_{||}$ on the axis.

***Remark:*** Once $U_{||}$ has been determined along different *t – coordinate lines (*for different *s* values), one may use the computed data to check for the constancy of $\frac{n}{B}$, $U_{||}$ and $\frac{nU_{||}}{B}$ on other field lines not on the axis.

***Evaluation of $U_\perp$, $U_\phi$, and n*:** Examination of Eqns. (30) and (33) reveals that the variables $U_\perp$ and $U_\phi$ are coupled and that these have to be solved simultaneously through step-by-step increments of *t* starting from *t* = 0. Eqns. (35), (37) and (38) take care of the initial values and the indeterminate terms at *t* = 0.

Next, one notes that Eq. (34) for *n*, is coupled to both $U_\perp$ and $U_\phi$, which have already been determined so that *n* may be found in a straightforward manner.

***Evaluation of E and Φ*: *E*** may be determined from Eqn. (10). Eqn. (7) with $f(t) \simeq 0$ yields $\Phi$. The constant of integration may be adjusted to match $\Phi$ with the experimentally measured plasma potential at a suitable point on the axis.

## 3. Magnetic coordinate system (MCS) and grid

Before going on to discussing the results, it is worth examining the nature of the MCS and its construction. In this section, only a few essential facts about the geometry and structure of the grid are presented. The details will be presented elsewhere.

Figures 2, 3 give an idea of the basic grid and the unit vectors ($\boldsymbol{e}_\perp$, $\boldsymbol{e}_{||}$) in the MCS. It can be seen that the MCS grid comprises discrete grid points in two dimensions, joining which gives rise to two sets of coordinates (*t*, *s*), where *t* measures distances along $\boldsymbol{e}_\perp$ (perpendicular to the field lines) and *s* measures distances along $\boldsymbol{e}_{||}$ (parallel to the field line); see discussion following Eq. (1). The spacing between grid points perpendicular to the field lines is $\Delta t$, while that between points parallel to the field line is $\Delta s$.

It is worth noting that the MCS grid construction has to take into account the spatial dependence of the magnetic field. *The latter gives rise to certain errors that are woven intimately within the process of mesh generation.* The bounds set on these errors determine the values $\Delta s$ / $\Delta t$ can have. The error is highest in regions where the field gradients are strong. In view of this, it is necessary to scan the entire region of interest. The scanning helps locate the regions that yield the largest error. *Conversely, for a given error it helps find the smallest value of* $\Delta s$ / $\Delta t$ *that meets the error criterion everywhere*. Too small an error bound yields very small step lengths that increase computation times significantly, rendering the computations impractical and expensive, while too large an error bound can compromise on accuracy. Therefore, an optimal choice of error bound that balances both requirements is necessary. Having arrived at a minimum step size, one needs to ensure that the step lengths $\Delta s$ or $\Delta t$ chosen or used for the calculations are smaller than the minimum step length.

Another aspect of MCS grid construction is the *determination of the coordinates of each grid point in the r–z plane*. Although this can be carried out in several ways, it turns out that adding grid points with fixed step lengths, $\Delta s$ or $\Delta t$ leads to a non-orthogonal (*t*, *s*) coordinate mesh, which is not acceptable. To overcome this an adaptive procedure was adopted in which $\Delta s$ / $\Delta t$ were treated as unknowns a priori, and their values determined self consistently for each point added to the $(t, s)$ mesh.

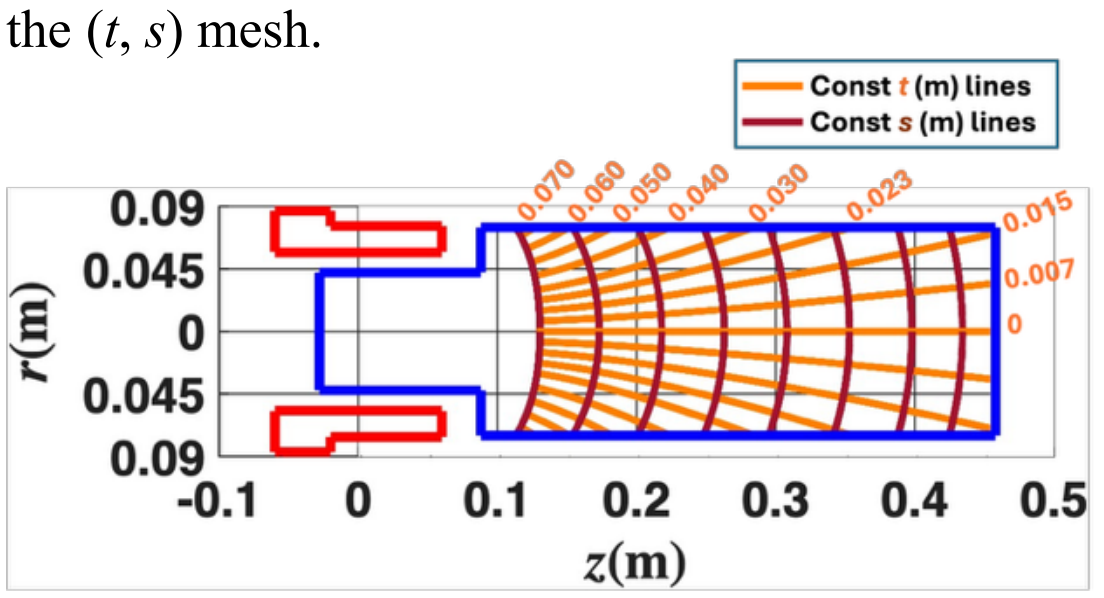

**Figure 5:** Maps of the $s$ = *constant contours* ($t$ – *coordinate; maroon*) and $t$ = *constant contours* ($s$ – *coordinate; orange*) in the $r$ – $z$ plane. The figure also shows the placement of the CEPS magnets and the expansion chamber. The $s$ and $t$ values can be read off at the ends. The strong bending of the $t$ – *coordinate* lines near the magnets indicate strong gradients of the magnetic field, getting gradually weaker with distance from the magnets. An important noticeable feature is that the extent of the $s$ - *coordinate* decreases sharply for $t > 1.5$ cm.

Figure 5 gives maps of the $s$ = *constant contours* ($t$ - *coordinate*) and $t$ = *constant contours* ($s$-*coordinate*) in the $r$–$z$ plane. Near the chamber end ($z \approx 45$ cm) and small $r$ ($\lesssim 2$ cm) the $s$-*coordinate* shows very little bending or curvature. On the other hand, close to the magnet corners ($z \approx 10$ - $15$ cm) and for large $r$ ($\approx 7$ cm), their bending becomes strong. In fact, the *range of the s – coordinate lines also decreases rapidly as the lines penetrate the chamber sidewall with increasing r and decreasing z*. A consequence of this is that the $t$ - *coordinate* also shows stronger bending at lower values of $z$, closer to the magnet.

In contrast to Figure 5, Figure 6 shows the $r$- and $z$-coordinate lines in the $t$–$s$ plane. It can be seen that the set of $r$ = *constant contours* are shaped in the form of a "funnel". This shape can be understood as follows. From Figure 5, one sees that *near the chamber end* where $z \approx s \approx 44$ cm, the full radial extent of the chamber (r ≈ 7 cm) is covered by a narrow range of $t \lesssim 1.5$ cm. On the other hand, closer to the magnets, $z \approx s \approx 10$ - $15$ cm, the full radial extent of the chamber (r ≈ 7 cm) extends up to $t \approx 7$ cm. It is this behaviour that is reflected in Figure 6, where near the chamber end, for $z \approx s \approx 47$ cm, the full radial extent of the chamber (r ≈ 7 cm) is compressed within $t < 2.0$ cm, whereas closer to the magnets for $z \approx s \approx 15$ cm, the full radial extent ($r \approx 7$ cm) requires a larger range of ($t \approx 7$ cm).

Figure 7 shows the allowed ranges of the $t$ – and $s$–*coordinates* for different $s$ = constant and $t$ = constant values, respectively, before the lines hit the chamber walls. In particular, one sees from Figure 7 (b), how the range of the $s$ – coordinate shrinks with increasing $t$. Thus, for large $t$, in plots with respect to $s$, the allowed range of $s$ will be small.

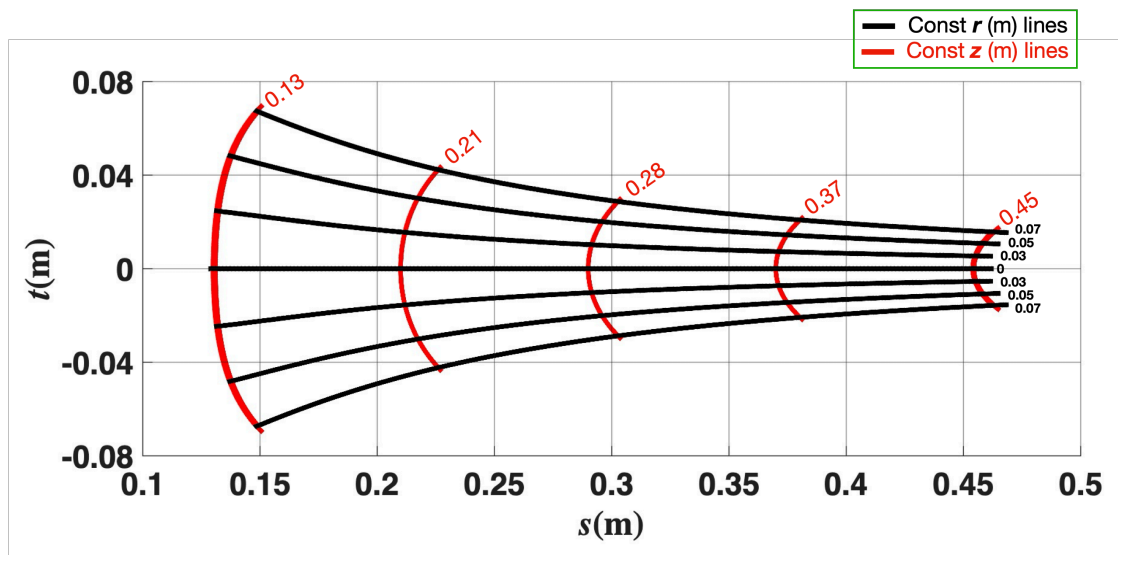

**Figure 6:** Maps of the $z$ = *constant contours (r – coordinate; red) and* $r$ = *constant contours* ($z$ – *coordinate*; black) in the $t$ – $s$ plane. The $r$ = *constant contours* are shaped in the form of a "funnel". See text for explanation.

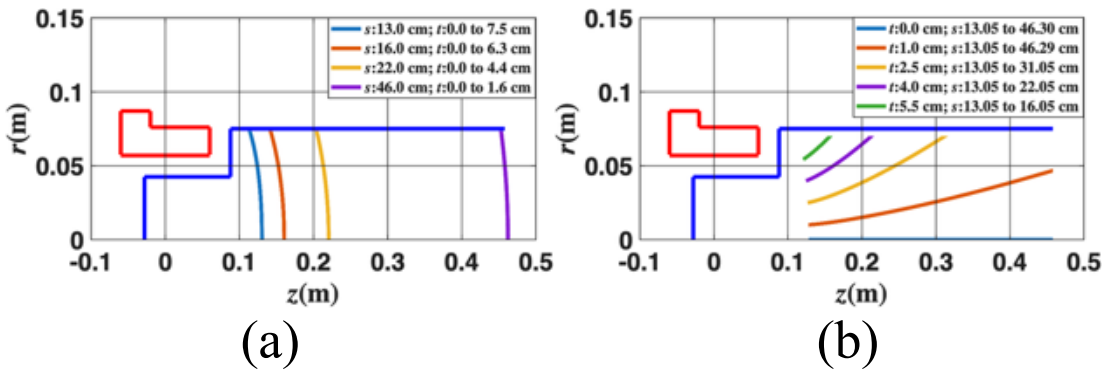

**Figure 7:** (a) Figure shows the allowed ranges of the $t$ – coordinate for different $s$ = constant values and (b) the $s$ coordinate for different $t$ = constant values in the $r$ – $z$ plane. Note how the allowed range of $s$ shrinks with increasing $t$.

## 4. Results and discussions

### 4.1 Validation of splitting of the ion equations

***Constancy of*** $\frac{nU_{||}}{B}$ ***on field lines***: Eqs. (41) or (42) *predict* the approximate constancy of $\frac{nU_{||}}{B}$ on other field lines as well, not just on the axis. While this prediction is a consequence of Eq. (31), the *actual solutions to be displayed below* are a direct consequence of Eqns. (27), (28), (30), (33) and (34) all of which are split equations, with the latter two being consequences of Eqns. (26) and (32). Therefore, the constancy of $\frac{nU_{||}}{B}$ on different field lines would be direct validation of the splitting method.

Since $\frac{nU_{||}}{B}$ is the product of $\frac{n}{B}$ and $U_{||}$, it is useful to examine the behaviour of these quantities along different field lines. Figs. 8 display normalized profiles of $n$ with respect to $s$ for $t$ = 0 - 4.0 cm, together with normalized profiles of $B$ for two pressures, ≈ 0.5 mTorr and ≈ 5 mTorr; $n$ and $B$ were normalized with respect to their values at the *same* point. Profiles of $\frac{n}{B}$ are also plotted alongside.

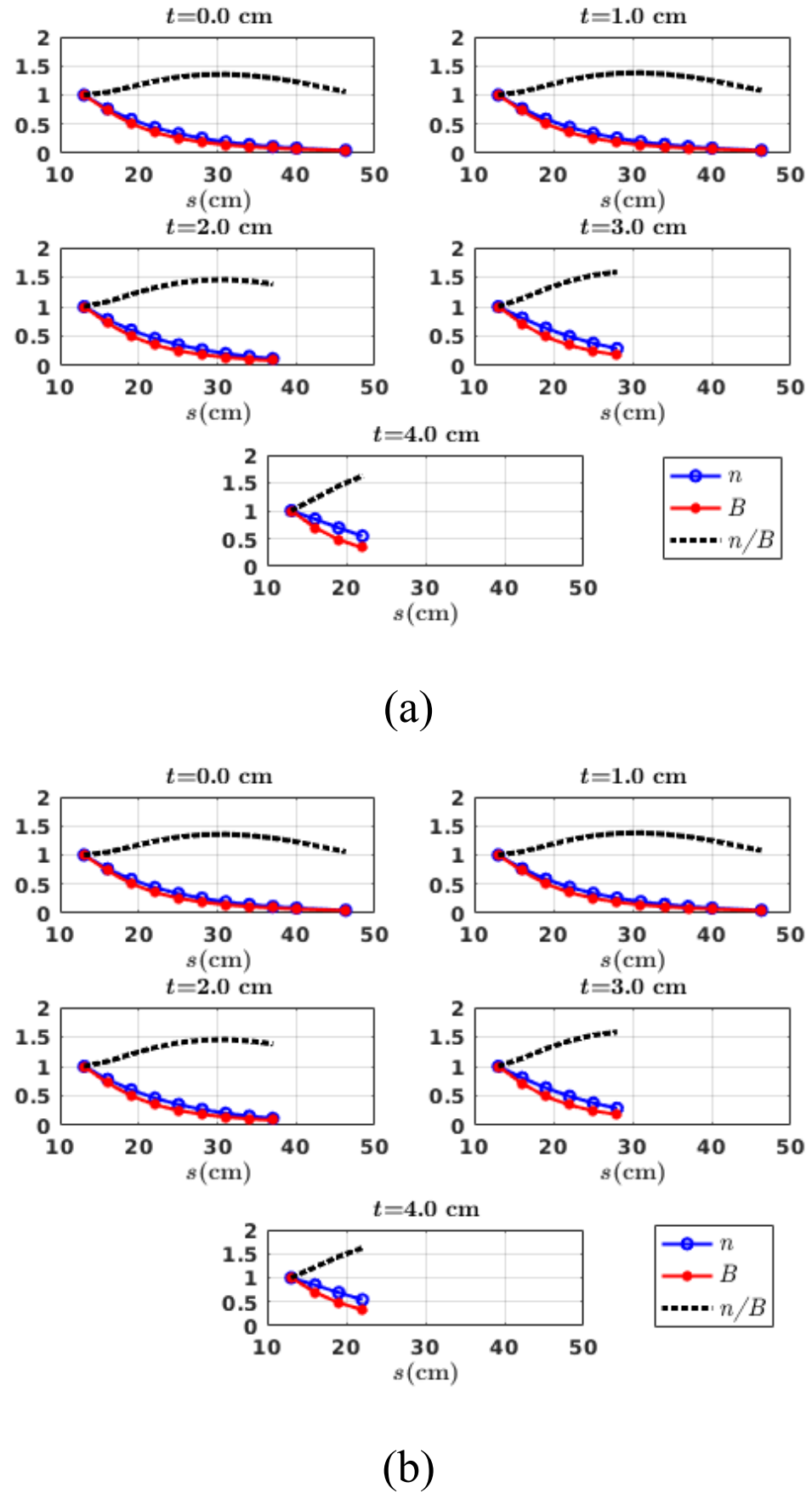

**Figure 8:** Profiles of $n$ (blue dot and line), $B$ (red dot and line) and $n/B$ (black, dashed) with respect to $s$ for different values of $t$ at two different pressures: (a) ≈ 0.5 mTorr and (b) ≈ 5.0 mTorr. The allowed range of $s$ decreases with increasing $t$. [See Fig. 7 and discussion at end of Sec. 3.]

For both pressures, one observes that the $n$ and $B$ profiles practically overlap for $t \leq 2.0$ cm. The upward bulges observed in the *central* portions of the profiles for $t$ = 0 and 1.0 are due to $n$ falling slightly less steeply than $B$ in the mid-section of the plot. This is also corroborated by the $\frac{n}{B}$ plot. To see the reason for these bulges, one notes that while the bulge at $t$ = 0 is due to deviation arising from the experimental data taken on the axis, the bulges for $t$ > 0 arise because the latter data was used for continuing the solutions for $t$ > 0. It may be recalled that $\frac{n}{B}$ scaling along field lines implies that the particles are bound or 'glued' to the field lines and so their density varies in proportion to $B$ (see Appendix B). However, ions obey this scaling because they are bound to the electrons through quasineutrality.

The overlap between the $n$ and $B$ profiles *reduces* as $t$ increases to ≈ 3 cm and ≈ 4 cm and, *$n$ falls less steeply than $B$, its profiles becoming flatter*. This increase can also be seen in the plots of $\frac{n}{B}$. It indicates a progressive decrease in the number of electrons bound to the field line.

The corresponding profiles for $U_{||}$ and $\frac{nU_{||}}{B}$ are shown in Figs. 9 and 10, respectively. It is seen that for $t$ = 0 - 2, where $\frac{n}{B}$ scaling holds well, $U_{||}$ is practically constant and so is $\frac{nU_{||}}{B}$, although the bulges in the $\frac{n}{B}$ profiles are conferred on them as well. *On the other hand, the increase in $\frac{n}{B}$ for $t$ = 3.0 and* 4.0 cm *is compensated by a corresponding decrease in $U_{||}$, so that $\frac{nU_{||}}{B}$ still remains reasonably constant.*

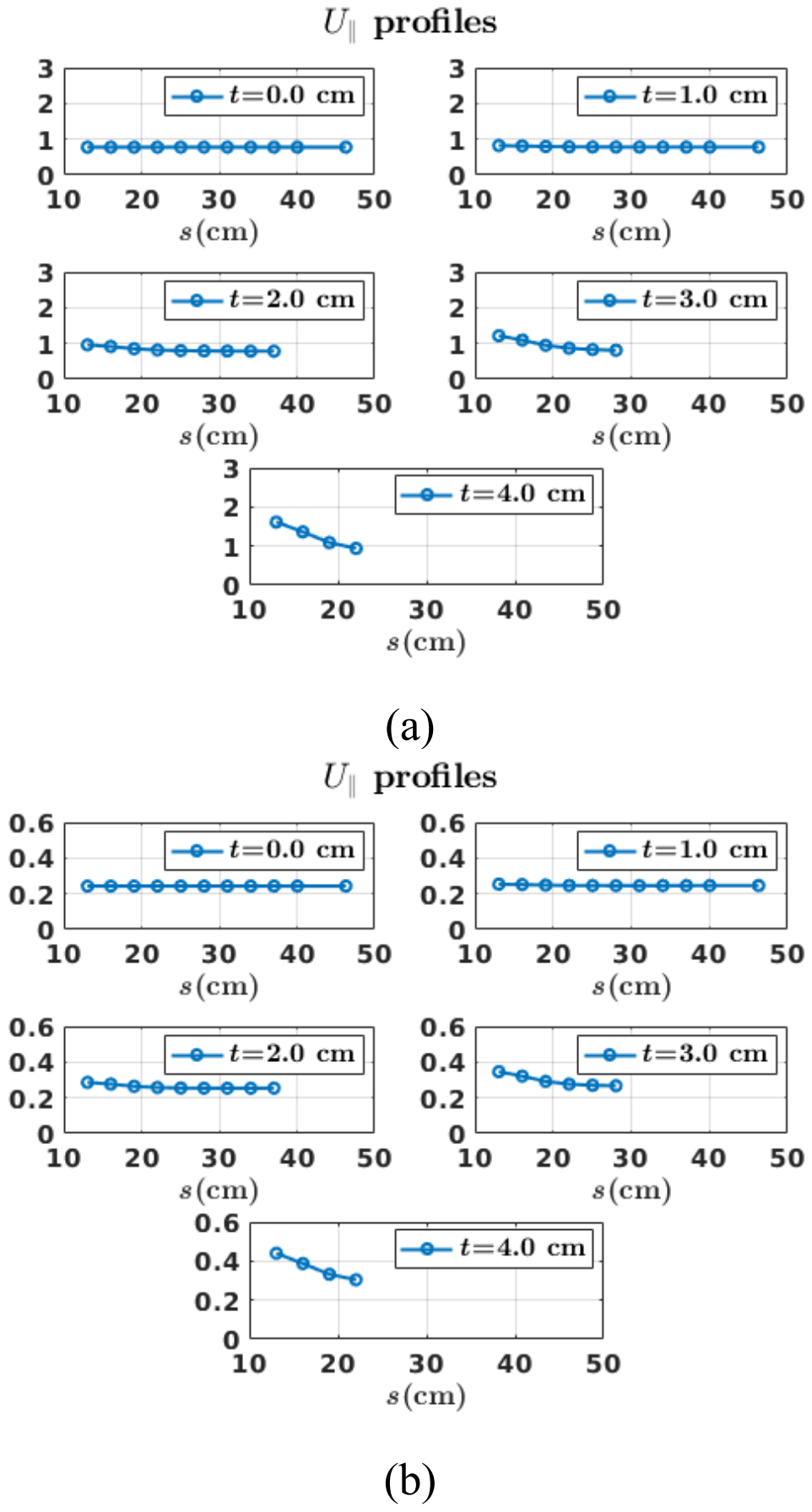

**Figure 9:** Similar plot as in Fig. 8 for $U_{||}$.

*In conclusion, one may say that $\frac{nU_{||}}{B}$ remains fairly constant along the field lines and that this feature is independent of whether $\frac{n}{B}$ and $U_{||}$ remain constant or vary with $s$. This confirms the validity of* Eq. (31) *as well as the splitting methodology in general, justifying its use for simplification of the ion flow and continuity equations*. It should be kept in mind however, that the detailed structure of the flow equations in the present work was derived keeping the electrons of the SVPS plasma in mind, as these exhibit unique properties. In the general case, this structure would be altered somewhat, although the methodology would still be valid.

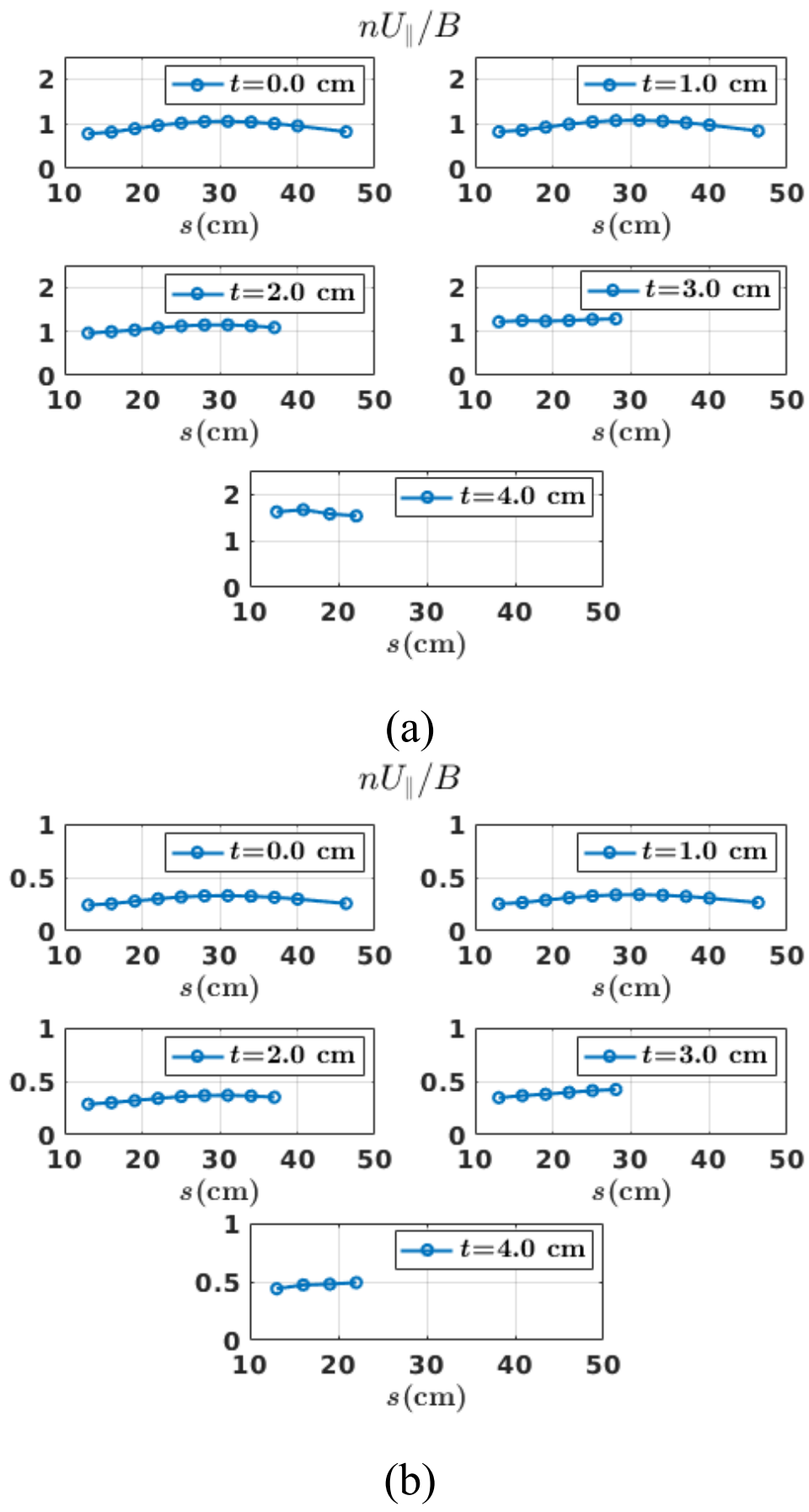

**Figure 10:** Similar plot as in Fig. 8 for ($nU\|/B$).

### 4.2 *n*, $U_\perp$ *and* $U_{||}$ profiles and plasma source characterization

***Characterization of plasma source – a Preview*:** An important aim of the present work is to obtain an approximate characterization of the plasma source using the results of the model. To this end, one has to use the computed *t*-profiles of the density and velocities, *closest to the source* ($s$ ≈13 cm). It may be recalled that the exit of the source was located at $z \approx s \approx 9$ cm and no experimental data was available in the region 9 cm ≤ $s$ < 13 cm. It was not possible to extend the current calculations for $s$ < 13 cm either, since the higher values of the magnetic field in that region (on account of close proximity to the magnets) required a finer MCS grid size than was used for the present calculations, for s > 13 cm. Investigations with a finer resolution of the grid to include the regions of strong field gradients are currently under consideration and will be presented elsewhere.

*Thus t-profiles of the density (n) and the velocities* ($U_\perp$ *and* $U_{||}$) *determined at s ≈13 cm* [see Fig. 7 (a)] *are taken as characteristics of the plasma source situated* ≈ 4 cm away, *at s* ≈ 9 cm. *In addition, the profiles can represent the plasma source only for* 0 ≤ $t$ ≤ 4 cm. This is because all field lines passing through the approximate coordinates, $s$ ≈ 13 cm, $t$ > 4 cm, *originate outside the source aperture* at $s$ ≈ 9 cm (see Fig. 7). It follows, therefore, that plasma on such lines must arrive there by cross-field transport and not directly from the source (to be discussed later).

***n profiles*:** Figs. 11(a – b) give normalized density profiles with respect to $t$ at two pressures, ≈ 0.5 mTorr and ≈ 5 mTorr. The profiles at both pressures are very similar; however, their normalizations are slightly different.

The $t$-profiles of the density show the steepest fall for $s$ ≈ 13 cm, while for higher values of $s$ (increasing distance from source mouth) the profiles become less steep, flattening out for $s$ ≥ 22 cm. [Once again, using Fig. 7(a), one notes that the larger values of $t$ ( ≈ 7 cm) are only accessible at low $s$ ( ≈ 13 cm), while the larger $s$ values ($s$ ≈ 46 cm) can occur only within a narrow range of $t$ (≤ 1.5 cm).]

As already stated above, the $t$-profiles of the density for $t \leq 4$ cm in Fig. 11 *closest to the source mouth* ($s$ ≈ 13 cm), *are taken to be characteristic of the density profiles and its values at the source exit*.

Fig. 12 gives a 2D color plot of the density in the $r$ – $z$ plane. It reproduces the features discussed above. However, the most noticeable feature of the plot is that it shows that the highest density (along $r$) occurs directly in front of the source aperture, falling away radially. This is truly remarkable, considering that the model only had on-axis values of the density as input (from the experiments), without any other information that could indicate the diameter of the source, situated ≈ 4 cm away from the first observation plane at $z \approx s \approx 13$ cm. In fact, one can see that highest density (bright yellow) is confined to a somewhat smaller zone (aperture). These features can also be seen clearly in the density profiles in the $r$-$z$ plane to be presented below in Figs. 19 – 20.

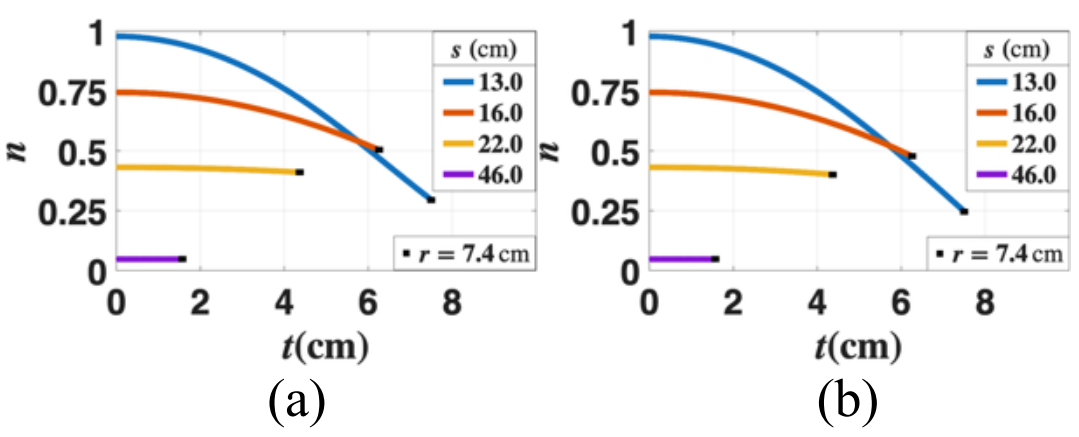

**Figure 11:** Profiles of normalized density ($n$) with respect to $t$ for different values of $s$ at two different pressures: (a) for ≈ 0.5 mTorr, and, (b) for ≈ 5.0 mTorr. The normalization density ≈ $1.1 \times 10^{18}$ m$^{-3}$ for 0.5 mTorr and $0.98 \times 10^{18}$ m$^{-3}$ for 5 mTorr. In these figures, the termination of the profiles on the chamber wall is indicated by the black marker.

As pointed out in Sec. 2.1, the density profiles with respect to $t$ (and $r$) would be affected due to ionization by the warm electrons. Comparison with experimentally measured radial density profiles (see Sec. 4.4) show that the profiles (shown with respect to $r$) are indeed affected, though not significantly.

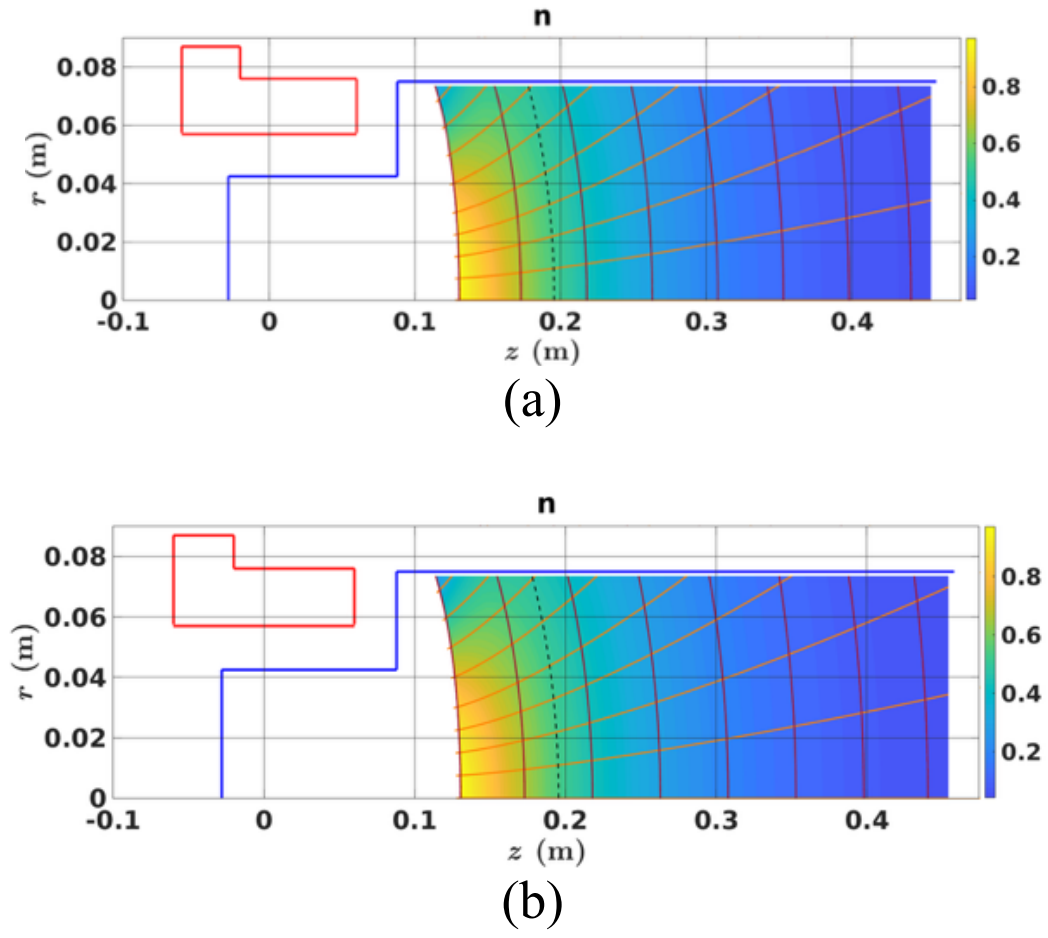

**Figure 12:** 2D Color profiles of $n$ in $r-z$ plane at two different pressures: (a) 0.5 mTorr and (b) 5.0 mTorr.

***$U_\perp$ and $U_{||}$ profiles*:** Figs. 13 (a, b) give profiles of $U_\perp$ with respect to $t$ at the two pressures, $\approx$ 0.5 and $\approx$ 5 mTorr; the normalizations are given in the figures. It is seen that the $U_\perp$ profiles at the two pressures are practically identical. Though small in magnitude, $U_\perp$ at $\approx$ 5 mTorr is about 5 times larger than at $\approx$ 0.5 mTorr.

It can be seen that the $t$ profiles of $U_\perp$ increase with $t$ up to about $t \approx 5$ cm. In particular, the $t$-profile for $s \approx 13$ cm, shows a steep rise up to $t \approx 7$ cm. As iterated before, *the t-profiles for $t \leq 4$ cm at $s \approx 13$ cm* (Figs. 13) *are taken to represent the $U_\perp$ characteristics of the source, both in terms of its profile, as well as in terms of its magnitude.*

The corresponding $t$-profiles of $U_{||}$ (Figs. 14) show marked increase with $t$ for low values of $s$ ($\leq$ 22 cm). Further away from the source, the profiles flatten. As before, the $t$-profile for $s \approx 13$ cm and $t \leq 4$ cm, may be taken to represent $U_{||}$ profile at the source exit.

It is clear that at low values of $t$ ($\leq$ 1.5 cm), $U_{||}$ is already at the drift speed at the source exit and hence, experiences very little change since the acceleration by the axial electric field cancels the retardation due to collisions. On the other hand, for larger $t$ ($t \geq 2.5$ cm) the ions exit the source with speeds (supersonic) far in excess of the drift speed and hence are first slowed down by collisions with the neutrals (within $\approx$10 – 15 cm), before settling down at the drift speed.

***Plasma source characteristics based on density and velocity profiles*:** Reviewing the density and velocity data, the two chief characteristics of the source emerge. Viewed from a direction facing the source (i.e. opposite the z direction), *the t-profile of the density is moderately convex while those of $U_{||}$ and $U_\perp$ profiles concave, with both being centered on the axis* (Figs. 11, 15 and 16). Of course, the $n$, $U_{||}$ and $U_\perp$ values at the source exit would be higher than those at $s$ $\approx$13 cm.

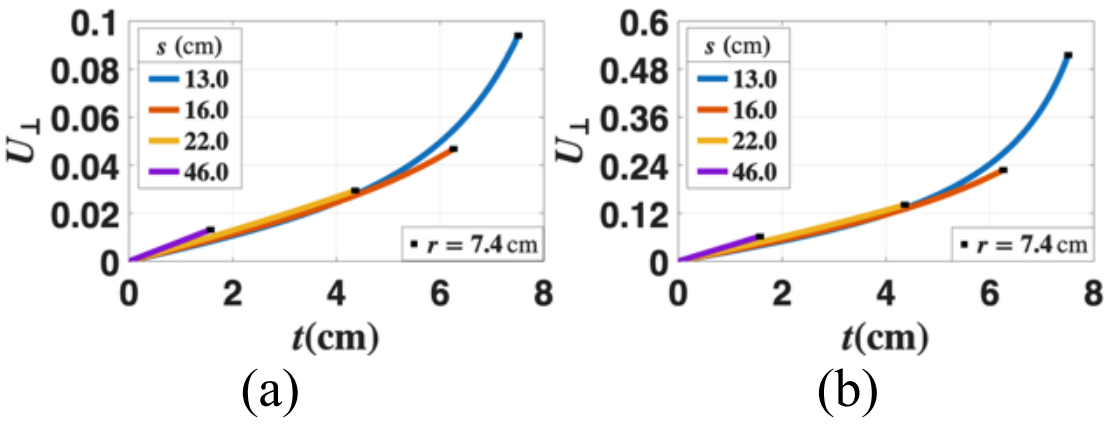

**Figure 13:** Profiles of $U_\perp$ with respect to $t$ for different values of $s$ respectively, at two different pressures: (a) for $\approx$ 0.5 mTorr, and, (b) for $\approx$ 5.0 mTorr. Their normalizations are: $\approx 2.63\times10^3$ m/s at $\approx$ 0.5 mTorr; and $\approx 2.47\times10^3$ m/s at $\approx$ 5 mTorr.

The corresponding $t$-profiles show marked increase with $t$ for low values of $s$ ($\leq$ 22 cm). Further away from the source, the profiles flatten. As before, the $t$-profile for $s \approx 13$ cm and $t \leq 4$ cm, may be taken to represent $U_{||}$ profile at the source exit.

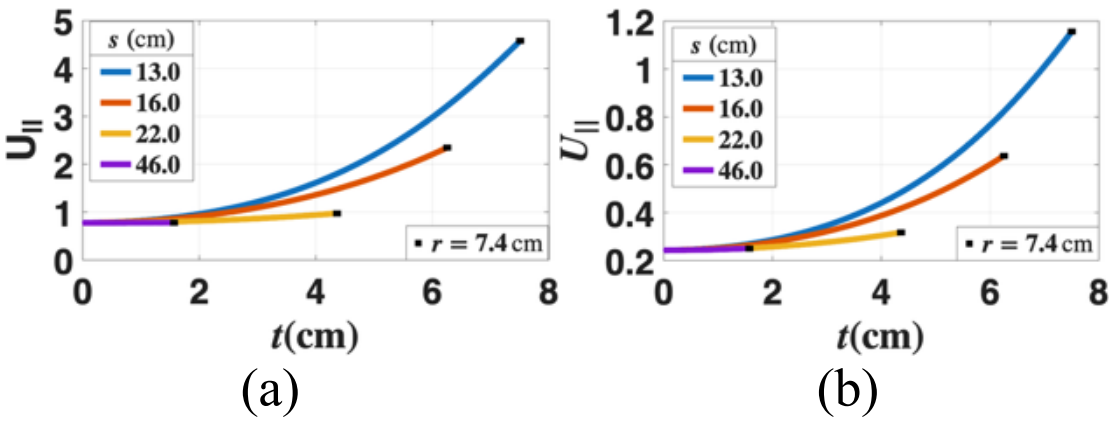

**Figure 14:** Similar plots as in Fig. 13, for $U||$. The normalizations are given in Fig. 13.

The above features indicate that *the microwave fields and steady state potential structures are such that at large radii they do not yield strong electron heating and plasma generation, although they produce considerable ion acceleration*. On the other hand, *close to the axis or in the paraxial region, electron heating is strong, leading to efficient ionization and plasma production, but the potential structures are weaker so that ion acceleration is modest*. To verify these features more measurements closer to the source mouth are needed. Although such measurements were not possible at the time the SVPS experiments were conducted, similar measurements undertaken recently using the CEPS in a larger system seem to confirm both these features of the CEPS.

### 4.3 Profiles in $r-z$ plane

***$U_r$ and $U_z$ profiles***: For completeness, some more profiles are presented. However, these are presented in the $r-z$ plane since they are more convenient from the viewpoint of visualization.

Figs. 15 and 16 give profiles of radial and axial profiles of $U_r$ for two pressures $\approx 0.5$ and $\approx 5$ mTorr. The plots reflect the basic features of $U_{||}$ and $U_{\perp}$ seen above. The radial profiles are concave (increasing outward) and the axial profiles falling monotonically because of friction. It must be kept in mind that apart from the variation with $t$ and $s$ or ($r$ and $z$) the $U_r$ profiles have added variation due to changing slope of the field lines with respect to $r$ and $z$. This complication is absent when using $U_{\perp}$ and $U_{||}$.

Likewise, radial, and axial profiles of $U_z$ are shown in Figs. 17 and 18, at $\approx 0.5$ and $\approx 5$ mTorr. The profiles are like those of $U_r$ as both are derived mainly from $U_{||}$ and mimic its prominent features.

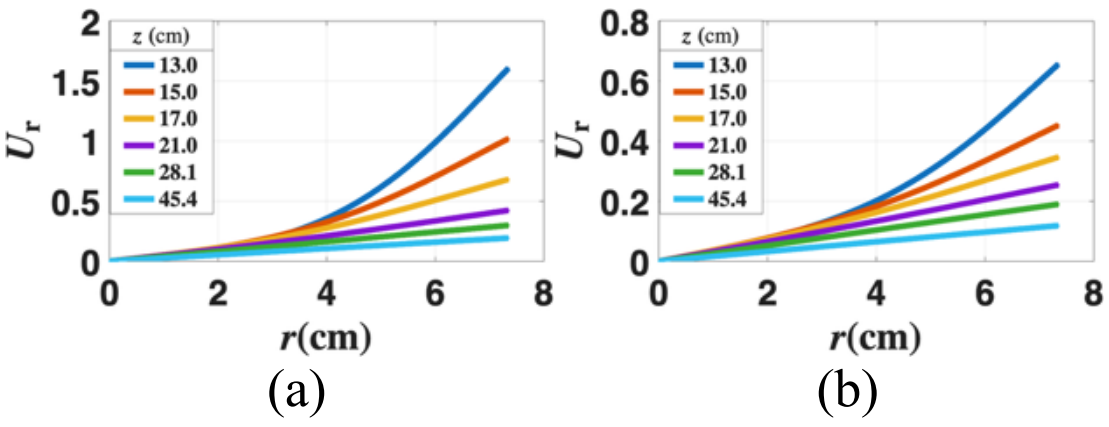

**Figure 15:** The radial velocity $U_r$ vs $r$ at different $z$ = *const*. planes; (a) at $\simeq 0.5$ mTorr and (b) at $\simeq 5.0$ mTorr

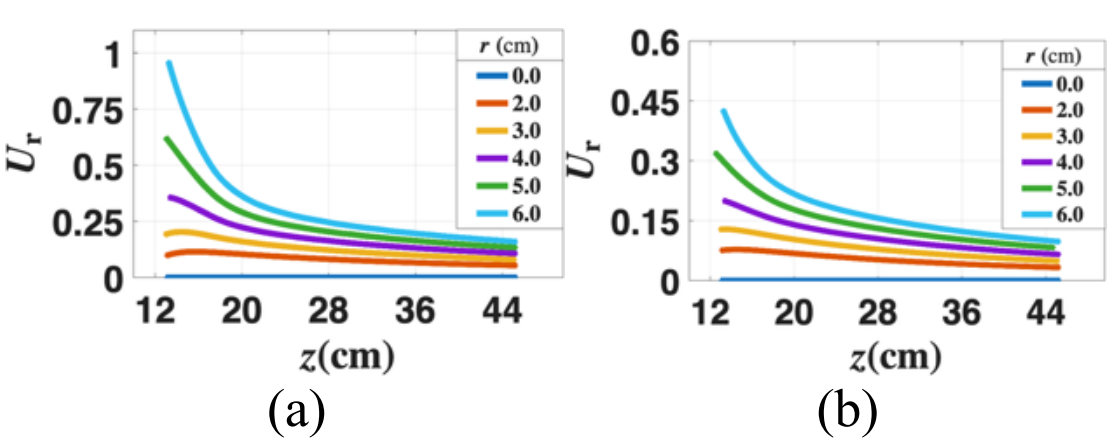

**Figure 16:** The radial velocity $U_z$ vs $z$ at different $r$ = *const*. planes; (a) at $\simeq 0.5$ mTorr and (b) at $\simeq 5.0$ mTorr

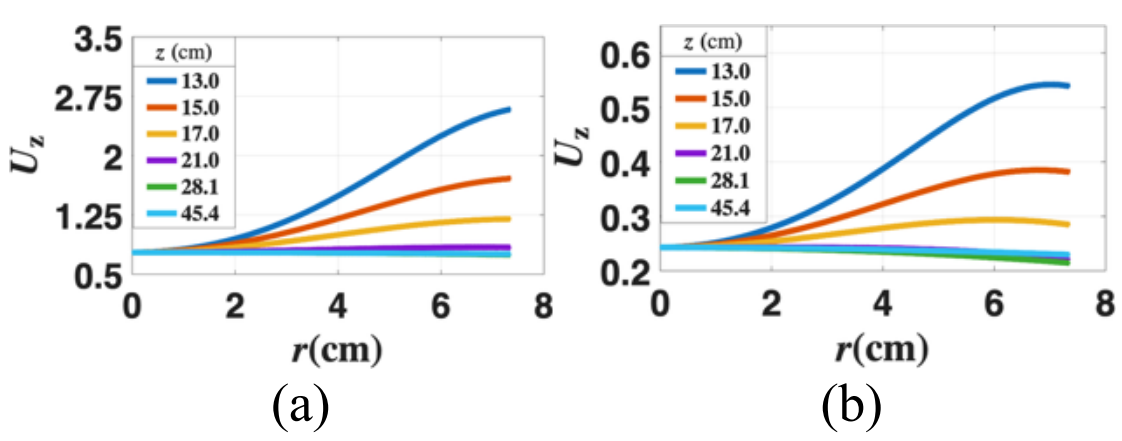

**Figure 17:** Similar plot as in Fig. 15 for the axial velocity $U_z$.

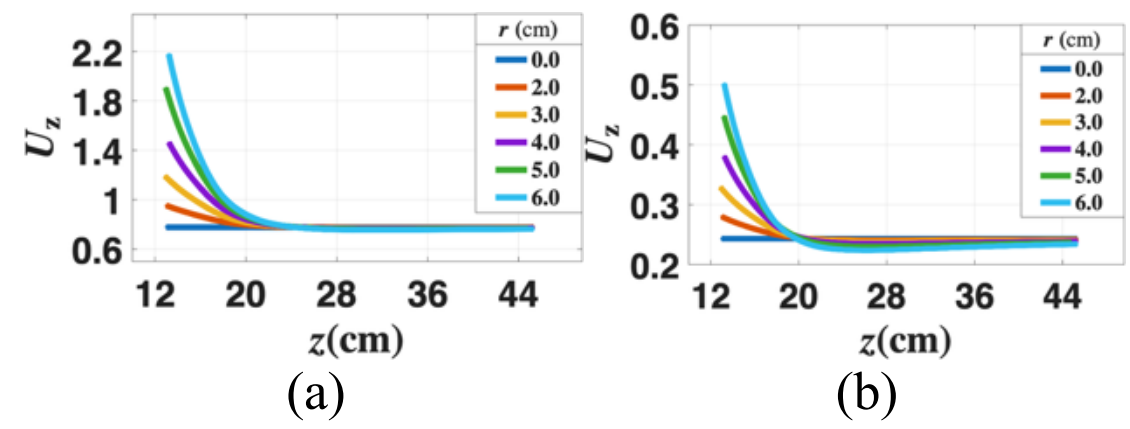

**Figure 18:** Similar plot as in Fig. 16 for the axial velocity $U_z$.

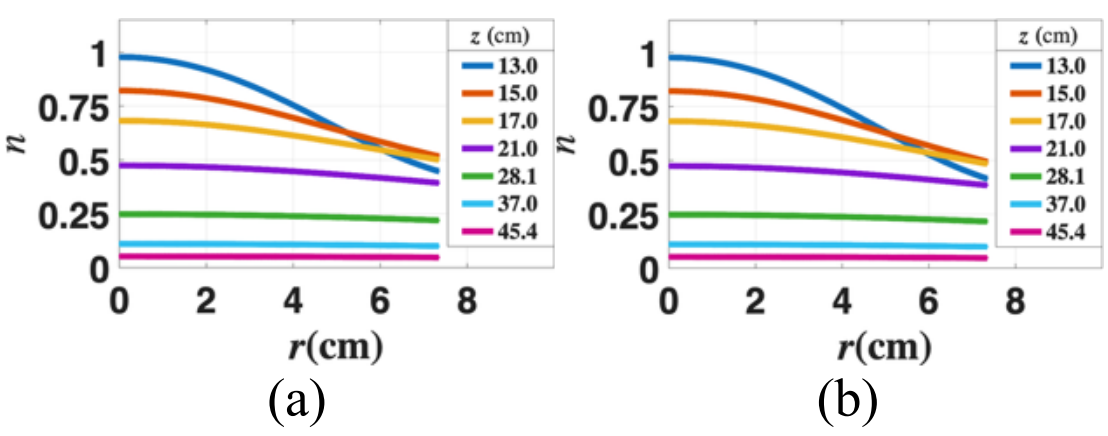

**Figure 19:** Radial profiles of $n$ at different $z$ = constant planes for two different pressures: (a) $\simeq 0.5$ mTorr and (b) $\simeq 5.0$ mTorr. Note that the density at z $\simeq$ 13 cm drops below those for $z \simeq 15$ and $\simeq 17$ cm for $r \gtrsim 5.5$cm giving a hump in the corresponding z-profile (see Fig. 20).

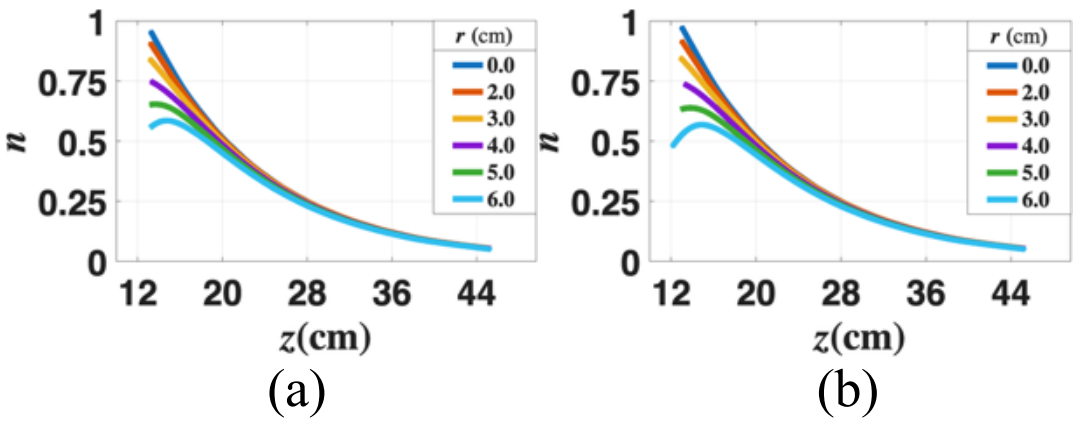

**Figure 20:** Profiles of $n$ vs $z$ for different values of $r$ at two different pressures: (a) $\simeq 0.5$ mTorr and (b) $\simeq 5.0$ mTorr. The slight hump in the profile for $r = 6$ cm between z $\simeq$ 13 cm and z $\simeq$ 17 cm is due the density at $z \simeq 13$ cm (see Fig. 19) falling below the densities for z $\simeq$ 15, 17 cm, etc., for r $\gtrsim$ 5.5cm.

***n profiles***: Figs. 19 and 20 give radial and axial profiles of the density in the $r-z$ plane at two different pressures, $\approx 0.5$ and $\approx 5$ mTorr. The profiles strongly resemble the profiles shown in Figs. 11. These have already been discussed in detail. The basic feature is that close to the source, although the peak central density is the highest, their profiles decrease fairly steeply. Away from the source, the densities are lower and the profiles quite flat. One notes that the radial profile at z $\approx$ 13 cm begins to fall at $r \approx 2$ cm. The fall is steeper than the profiles at z $\approx$ 15 and 17 cm, falling below those profiles. The latter behaviour shows up in the axial plots as well. For example, in Fig. 20 (a) one notes that the profile for $r \approx 6$ cm shows a small peak between z $\approx$ 13 and 17 cm. To understand the latter, one may compare with the radial profiles in Fig. 19 (a) at a fixed radius, $r \approx 6$ cm. At this radius, looking at values of $n$ for z ≈ 13, 15 and 17 cm, one finds the lowest density at z ≈ 13 cm; at z ≈ 15 cm one finds $n$ increasing a little and then again falling at $z$ ≈ 17 cm. This is precisely the behaviour that is seen in the axial plot Fig. 20 (a) between z ≈ 13 and 17 cm.

### 4.4 Comparison with experiment

The SVPS had only one radial port located at $z$ = 28.1 cm. The radial experimental data are displayed in Figs. 21 [47]. Figures (a), (b) and (c) give profiles for ≈ 0.5 mTorr and Figs. (d), (e) and (f), for ≈ 5 mTorr. Figs. (a) and (d) give the density plots: These include the experimentally obtained warm and bulk electron densities, $n_w$ and $n_b$. Also included are the profiles for theoretical density $n$, already presented in Figs. 19 (a) and (b). Analogously, plots of the experimental warm and bulk electron temperatures ($T_{we}$, $T_{be}$) and the corresponding theoretical electron temperature $T_e$ (= constant in SVPS) are shown in Figs. (b) and (e). Finally, the experimental and theoretical plots of the plasma potential, $\Phi_b$ and $\Phi$ are shown in the Figs. (c) and (f).

From Figures 21 (c) and (f) one notes that both the experimental and theoretical potential profiles are practically constant and overlap, giving a vanishing $E_r$ and a constant $n$ along $r$ since in the SVPS model here $\Phi = T_e \ln(n)$ and $E_r = -\partial\Phi/\partial r$, by definition. Figures (b) and (e) show that $T_{be}$ is constant for both pressures as is the theoretically assumed $T_e$. $T_{we}$, is close to ≈ 100 eV, and is also practically constant at both pressures, although it exhibits a slight hump at $r$ ≈ 2 cm at the lower pressure.

The theoretical density plot ($n$) is constant as predicted by the potential data in Figures (c) and (f). As stated earlier, for both pressures, the warm density, $n_w$ is very small (≈ 0.1% $n_b$) and cannot play any significant role in the dynamics of the plasma [47,48], although it can ionize efficiently as $T_{we}$ ≈ 100 eV. It may be noted however, that on account of the $n/B$, scaling on the axis, ionization by the warm electrons *does not affect the axial profiles*. It affects the radial profiles though, as seen from Figs. (a) and (c).

It is seen that $n_w$ falls almost linearly implying that the electrons added by it (via ionization) will also have a similar profile. At this point one must distinguish between *two* types of low temperature electrons according to the SVPS experiment: (i) The *flow electrons* with density $n_F$ that arrive at the given plane by flow. Although these include electrons added by ionization by the warm population, their radial profile is rendered flat by the constant potential in the plane and the relation, $\Phi = T_e \ln(n)$. (ii) The second type of electrons in the given plane are those that are added by ionization by the warm population in the plane. Their density $n_I$ will have the profile of $n_w$. On the other hand, the flow electrons will correspond to the theoretical electron population and will have a flat radial profile density (without the inclusion of the ionization by the warm electrons).Thus, adding $n_I$ to $n_F$ should produce a *radially decreasing population with density*, $n_T$ (= $n_I$ + $n_F$) *just as seen in the experiments*.

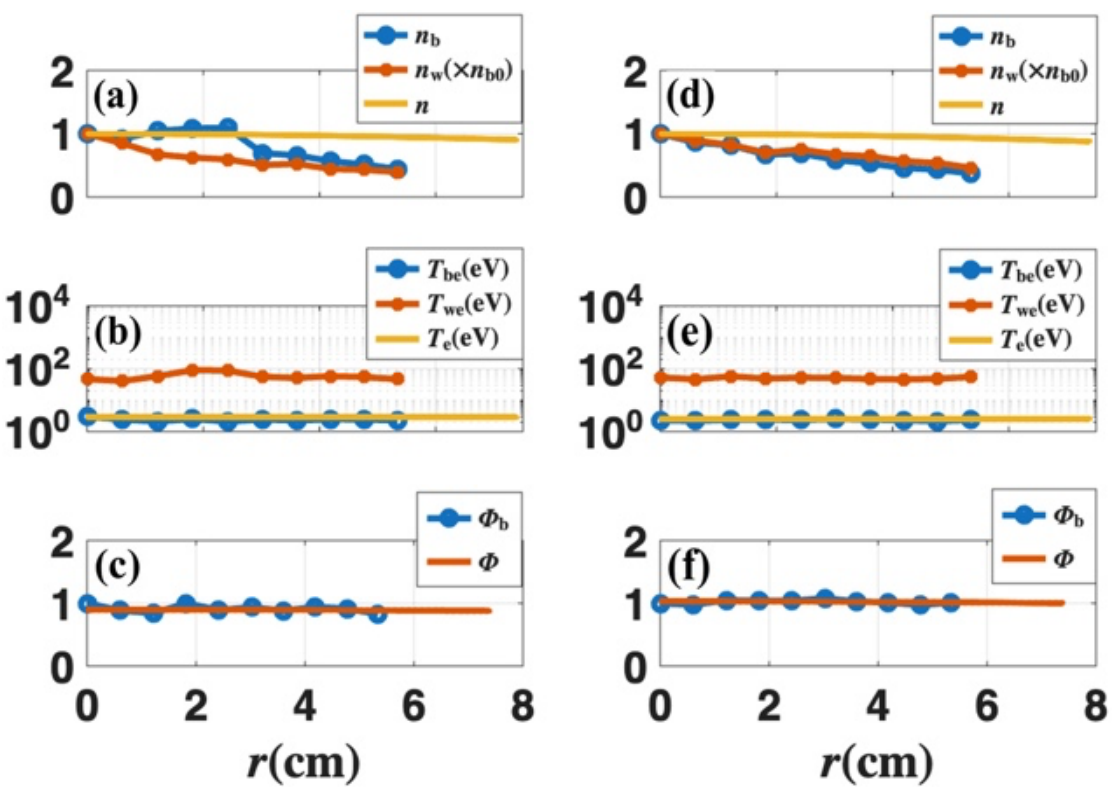

**Figure 21:** Comparison of radial profiles of the flow model and the experimental measurements for 0.5 mTorr (a, b, c) and 5 mTorr (d, e, f) pressure cases at $z$ = 0.281 m constant line. $n_b$ and $n_w$ are the experimentally measured normailized densities of bulk and warm electron populations, whereas $n$ is the normalized density from the flow model. Similarly, the corresponding temperatures ($T_{be}$, $T_{we}$, $T_e$) and potentials ($\Phi_b$, $\Phi$.) have been shown.

Examining the data in Fig 21 (d) for 5 mTorr, one sees that the normalized profile of $n_b$ is the same as that of normalized, $n_w$. This is precisely what was predicted above for the profile of $n_T$ (that it will track $n_w$ approximately). Fig. 21 (a) also shows a similar trend except for a higher density at low radii owing to a hump in $T_{we}$ as seen from Fig. 21 (b). The latter hump in $T_{we}$ enhances ionization locally, resulting a slightly altered profile for $n_b$.

## 5. Conclusions:

In this paper, a fluid model for expanding magnetised plasma in the magnetic coordinate system has been investigated, and based on the ordering of the flow variables, a new methodology, i.e., the splitting of the flow equations, has been presented for modelling the experimental plasma flow from the ECR source. Due to limitations in experimental measurements near the source, the on-axis flow measurement has been used as the initial input for the model's flow variables. Also, by establishing the experimentally measured $n/B$ scaling theoretically on the axis, the flow variables have been solved in the off-axis region in magnetic coordinate systems. Later, the model results were compared with the only available experimental results. It is predicted that the warm population plays an important role inspite of not affecting the plasma flow and must be considered in the future model updates. Besides this, the model is more general, with a strong interest in both academic and industry domains.

## APPENDIX - A

### Formulas in Magnetic Coordinate System and Transformations

***Forward transformations:*** The magnetic field in this work is that of the ring magnets placed coaxially around the plasma source section, collinearly with the axis of the expansion chamber. It produces an axisymmetric field about the $z$–axis. In cylindrical coordinates ($r$, $\varphi$, $z$) with the $r$ = 0 axis coinciding with the $z$–axis, axisymmetry implies the field would be independent of $\varphi$ and that *the field lines have no twist and lie in constant $\varphi$ or r - z planes* ($\varphi$: 0 → 360° ). The symmetry also implies that it suffices to consider the field lines in a *single* $r$ – $z$ plane rendering the $\varphi$ coordinate irrelevant; thus, the problem reduces to 2D with just two coordinates, ($r$, $z$).

In view of the above, the magnetic coordinate system (MCS) will also have two coordinates. These are designated ($t$, $s$), where $t$ is the coordinate *perpendicular* to the magnetic field lines (in a fixed $r$ - $z$ plane with constant $\varphi$) and $s$ is the coordinate *along* the field lines. In 3D, the cylindrical coordinates ($r$, $\varphi$, $z$) may be replaced by the MCS coordinates, ($t$, $\varphi$, $s$), the $\varphi$ coordinate being identical in both coordinate systems. This Appendix gives transformation formulas between cylindrical coordinates ($r$, $\varphi$, $z$) and magnetic coordinates ($t$, $\varphi$, $s$) in addition to other relevant formulas needed in the main paper.

The unit vector $\boldsymbol{e}_{||}$ is tangent to the field line and changes direction with the field line. It is defined through $\boldsymbol{B}$ as follows.

$$\boldsymbol{B} = B\,\boldsymbol{e}_{||} = B_r\,\boldsymbol{e}_r + B_z\,\boldsymbol{e}_z$$
$$\Rightarrow \boldsymbol{e}_{||} = \frac{B_r}{B}\boldsymbol{e}_r + \frac{B_z}{B}\boldsymbol{e}_z \qquad \text{(A-1)}$$

$B_r$ and $B_z$ are the $r$ – and and $z$–components of the magnetic field, $B = [B_r^2 + B_z^2]^{\frac{1}{2}}$ . $\boldsymbol{e}_r$ and $\boldsymbol{e}_z$ are the unit vectors along radial and z-directions, respectively. The unit vectors, ($\boldsymbol{e}_r$, $\boldsymbol{e}_z$) form a 2D basis in the $r$ - $z$ plane such that the vector product, $\boldsymbol{e}_r \times \boldsymbol{e}_z$ points *into* the $r$ - $z$ plane. The unit vector, $\boldsymbol{e}_\perp$ is obtained by rotating $\boldsymbol{e}_{||}$ through 90° so that the unit vectors ($\boldsymbol{e}_\perp$, $\boldsymbol{e}_{||}$) also form a 2D basis in the $r$ – $z$ or $t$ – $s$ plane with the product, $\boldsymbol{e}_\perp \times \boldsymbol{e}_{||}$ pointing *into* the $r$ - $z$ plane. Defining $b_r = B_r/B$ , $b_z = B_z/B$ gives,

$$\boldsymbol{e}_\perp = b_z\,\boldsymbol{e}_r - b_r\,\boldsymbol{e}_z;\ \boldsymbol{e}_{||} = b_r\,\boldsymbol{e}_r + b_z\,\boldsymbol{e}_z \qquad \text{(A-2)}$$

Figure 2 shows the ($t$, $s$) grid in the ($r$, $z$) plane, and the unit vectors ($\boldsymbol{e}_\perp$, $\boldsymbol{e}_{||}$) in the MCS are shown in Figure 4. Let $\Delta\boldsymbol{R} = \boldsymbol{R}_2 - \boldsymbol{R}_1$, be an infinitesimal change in the position vector in a translation from point $\boldsymbol{R}_1 \rightarrow \boldsymbol{R}_2$.

Case 1: Let $\Delta\boldsymbol{R}$ be a displacement $\Delta t$, along $\boldsymbol{e}_\perp$. Using (A-2) gives

$$\Delta\boldsymbol{R} = \Delta t\ \boldsymbol{e}_\perp = \Delta t\ (b_z\ \boldsymbol{e}_r - b_r\ \boldsymbol{e}_z) \Rightarrow \Delta r\ \boldsymbol{e}_r + \Delta z\ \boldsymbol{e}_z$$
$$\Rightarrow \Delta r = \Delta t\ b_z,\ \Delta z = -\,\Delta t\ b_r$$
$$\Rightarrow \frac{\partial r}{\partial t}\Big|_s = b_z,\ \frac{\partial z}{\partial t}\Big|_s = -b_r \qquad \text{(A-3)}$$

Case 2: Likewise, if $\Delta\boldsymbol{R} = \Delta s\ \boldsymbol{e}_{||}$ (a displacement $\Delta s$, along $\boldsymbol{e}_{||}$), one can show using (A-1) and (A-2)

$$\Rightarrow \qquad \frac{\partial r}{\partial s}\Big|_t = b_r, \qquad \frac{\partial z}{\partial s}\Big|_t = b_z \qquad \text{(A-4)}$$

The partial derivative *operators* in magnetic coordinates can thus be written as

$$\frac{\partial}{\partial t}\Big|_s = b_z\frac{\partial}{\partial r} - b_r\frac{\partial}{\partial z}$$
$$\frac{\partial}{\partial s}\Big|_t = b_r\frac{\partial}{\partial r} + b_z\frac{\partial}{\partial z} \qquad \text{(A-5)}$$

Using (A-3) and (A-4) one may write the *differential* relations,

$$dr = b_z dt + b_r ds,$$
$$dz = -b_r dt + b_z ds \qquad \text{(A-6)}$$

***Inverse transformations:*** Likewise, considering infinitesimal displacements along $\boldsymbol{e}_r$ and $\boldsymbol{e}_z$ one may derive the inverse relations analogous to (A-3) and (A-4),

$$\frac{\partial t}{\partial r}\Big|_z = b_z, \qquad \frac{\partial t}{\partial z}\Big|_s = -b_r \qquad \text{and}$$
$$\frac{\partial s}{\partial r}\Big|_t = b_r, \qquad \frac{\partial s}{\partial z}\Big|_t = b_z \qquad \text{(A-7)}$$

Using (A-7) gives the relations

$$\frac{\partial}{\partial r}\Big|_z = b_z\frac{\partial}{\partial t} + b_r\frac{\partial}{\partial s}$$
$$\text{and} \qquad \frac{\partial}{\partial z}\Big|_r = -\,b_r\frac{\partial}{\partial t} + b_z\frac{\partial}{\partial s} \qquad \text{(A-8)}$$

$$dt = b_z dr - b_r dz,$$
$$\text{and} \qquad ds = b_r dr + b_z dz \qquad \text{(A-9)}$$

***Derivatives of the unit vectors:*** One also needs the derivatives of the unit vectors $\boldsymbol{e}_\perp$ and $\boldsymbol{e}_{||}$ with respect to $t$ and $s$. Differentiating $\boldsymbol{e}_\perp$ in (A-2) with respect to $t$ and noting that the $\frac{\partial}{\partial t}$ operator [given by the 1st of

(A-5)] will not operate on the unit vectors $\boldsymbol{e}_{\rm r}$ and $\boldsymbol{e}_{\rm z}$, one obtains

$$\frac{\partial \boldsymbol{e}_{\perp}}{\partial t} = \frac{1}{b_{\rm r}}\frac{\partial b_{\rm z}}{\partial t}\boldsymbol{e}_{||}$$

Following through the above procedure, one may collect the results to write,

$$\frac{\partial \boldsymbol{e}_{\perp}}{\partial t} = a_1\boldsymbol{e}_{||}, \qquad \frac{\partial \boldsymbol{e}_{\perp}}{\partial s} = a_2\boldsymbol{e}_{||}, \qquad \frac{\partial \boldsymbol{e}_{||}}{\partial t} = -a_1\boldsymbol{e}_{\perp} \quad \text{and} \quad \frac{\partial \boldsymbol{e}_{||}}{\partial t} = a_2\boldsymbol{e}_{\perp} \qquad \text{(A-10)}$$

In the above,

$$a_1 = \frac{1}{b_{\rm r}}\frac{\partial b_{\rm z}}{\partial t} = -\frac{1}{b_{\rm z}}\frac{\partial b_{\rm r}}{\partial t}, \qquad a_2 = \frac{1}{b_{\rm r}}\frac{\partial b_{\rm z}}{\partial s} = -\frac{1}{b_{\rm z}}\frac{\partial b_{\rm r}}{\partial s} \qquad \text{(A-11)}$$

One also needs the derivatives with respect to $\varphi$

$$\frac{\partial \boldsymbol{e}_{\perp}}{\partial \phi} = b_{\rm z}\boldsymbol{e}_{\varphi} \quad \text{and} \quad \frac{\partial \boldsymbol{e}_{||}}{\partial \phi} = b_{\rm r}\boldsymbol{e}_{\varphi} \qquad \text{(A-12)}$$

***The $\nabla$ operator in magnetic coordinates and related formulas:*** One has,

$$\nabla = \boldsymbol{e}_{\rm r}\frac{\partial}{\partial r} + \frac{\boldsymbol{e}_{\phi}}{r}\frac{\partial}{\partial \phi} + \boldsymbol{e}_{\rm z}\frac{\partial}{\partial z} = \boldsymbol{e}_{\perp}\frac{\partial}{\partial t} + \frac{\boldsymbol{e}_{\phi}}{r}\frac{\partial}{\partial \phi} + \boldsymbol{e}_{||}\frac{\partial}{\partial s} \qquad \text{(A-13)}$$

***$\nabla . \boldsymbol{B}$ and $\nabla \times \boldsymbol{B}$ in (t, s) coordinates:*** The expressions for divergence and curl will be

$$\nabla \cdot \boldsymbol{B} = \frac{\partial B_{\perp}}{\partial t} + \frac{\partial B_{||}}{\partial s} + \left[\frac{b_{\rm r}}{r} - a_1\right]B_{||} + \left[\frac{b_{\rm z}}{r} + a_2\right]B_{\perp}$$

In MCS, $B_{||} = B$ and $B_{\perp} = 0$, so that

$$\nabla \cdot \boldsymbol{B} = \frac{\partial B}{\partial s} + \left[\frac{b_{\rm r}}{r} - a_1\right]B = 0$$

$$\Rightarrow \qquad a_1 = \frac{b_{\rm r}}{r} + \frac{\partial \ln B}{\partial s} \qquad \text{(A-14)}$$

And,

$$\nabla \times \boldsymbol{B} = \left[\boldsymbol{e}_{\perp}\frac{\partial}{\partial t} + \frac{\boldsymbol{e}_{\phi}}{r}\frac{\partial}{\partial \phi} + \boldsymbol{e}_{||}\frac{\partial}{\partial s}\right] \times B\boldsymbol{e}_{||} = -\left[\frac{\partial B}{\partial t} + a_2 B\right]\boldsymbol{e}_{\phi} = 0$$

$$\Rightarrow \qquad a_2 = -\frac{\partial \ln B}{\partial t} \qquad \text{(A-15)}$$

***$(\boldsymbol{v} \cdot \nabla)\boldsymbol{v}$ in (t, s) coordinates:*** Using the $\nabla$ operator in (A-13), one has

$$(\boldsymbol{v} \cdot \nabla)\boldsymbol{v} = (\boldsymbol{e}_{\perp}v_{\perp} + \boldsymbol{e}_{\phi}v_{\phi} + \boldsymbol{e}_{||}v_{||}) \cdot (\boldsymbol{e}_{\perp}\frac{\partial}{\partial t} + \frac{\boldsymbol{e}_{\phi}}{r}\frac{\partial}{\partial \phi} + \boldsymbol{e}_{||}\frac{\partial}{\partial s})(\boldsymbol{e}_{\perp}v_{\perp} + \boldsymbol{e}_{\phi}v_{\phi} + \boldsymbol{e}_{||}v_{||})$$

Using azimuthal symmetry, the $(\boldsymbol{v} \cdot \nabla)\boldsymbol{v}$ term becomes,

$$(\boldsymbol{v} \cdot \nabla)\boldsymbol{v} = \boldsymbol{e}_{\perp}\left[v_{\perp}\frac{\partial v_{\perp}}{\partial t} - a_1 v_{\perp}v_{||} - b_{\rm z}\frac{v_{\phi}^2}{r} + v_{\perp}\frac{\partial v_{||}}{\partial s} - a_2 v_{||}^2\right] + \boldsymbol{e}_{\phi}\left[v_{\perp}\frac{\partial v_{\phi}}{\partial t} + b_{\rm z}\frac{v_{\phi}v_{\perp}}{r} + b_r\frac{v_{\phi}v_{||}}{r} + v_{||}\frac{\partial v_{\phi}}{\partial s}\right] + \boldsymbol{e}_{||}\left[v_{\perp}\frac{\partial v_{||}}{\partial t} + a_1 v_{\perp}v_{||} - b_r\frac{v_{\phi}^2}{r} + v_{||}\frac{\partial v_{||}}{\partial s} + a_1 v_{\perp}^2\right] \qquad \text{(A-16)}$$

***$a_1$ and $a_2$ on the axis (t = 0, s):*** Using $B_{\rm r} = 0$ on the axis, gives,

$$a_1(t = 0, s) = \frac{1}{2B}\frac{\partial B}{\partial z}; \quad a_2(t = 0, s) = -\frac{1}{B}\frac{\partial B_r}{\partial z} = 0 \qquad \text{(A-17)}$$

## APPENDIX - B

### Physical Significance of *n*/*B* Scaling

It will be shown below that *n*/*B* scaling is intimately connected to electrons being strongly bound or 'glued' to the magnetic field lines. In fact, it will be seen below that one needs *constant electron temperature* $T_{\rm e}$ for *n*/*B* scaling to hold. A still stronger result [48] can be deduced if *adiabatic constancy of the magnetic moment* in a magnetic field varying slowly in space is invoked, in which case the electrons obey the *double adiabatic equation of state*.

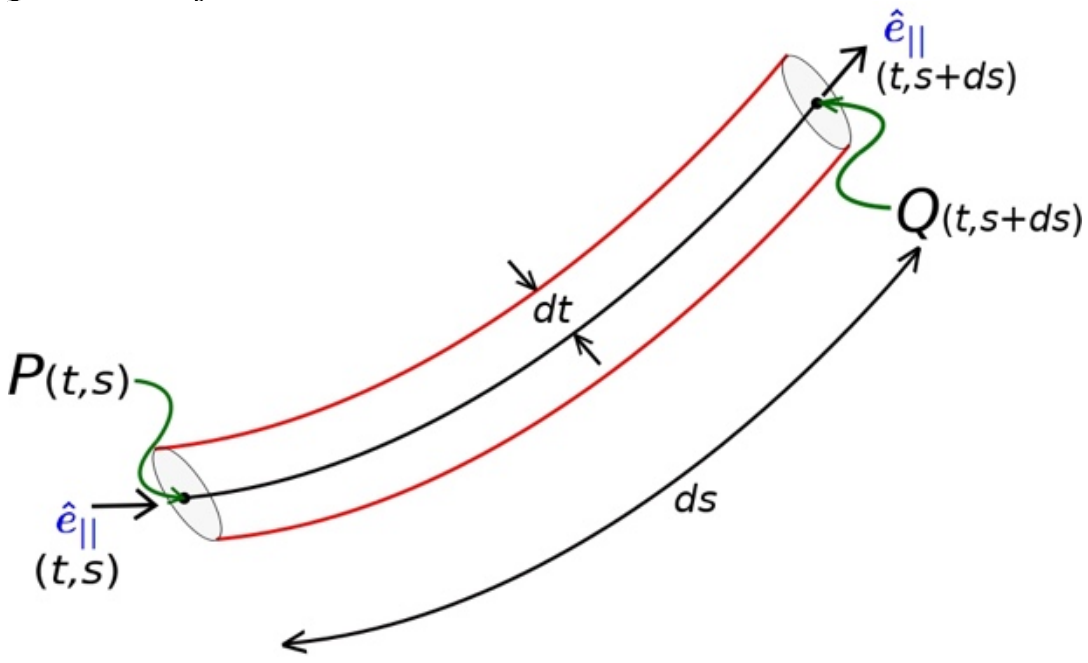

Figure B-1: Schematic of a thin tube around a field line

***Magnetic flux through a thin tube centred on a field line:*** Consider, as shown in Figure B-1, an infinitesimally thin tube of radius d*t*, centred on a field line between points P (*t*, *s*) and Q (*t*, *s* + d*s*). *The tube is not a flux tube, so that field lines will intersect the tube sidewalls along its length.*

The *magnetic flux entering* the tube at point P (*t*, *s*) is

$$\psi_{\rm in}(t, s) = \boldsymbol{B}(t, s) \cdot d\boldsymbol{A}(t, s) \qquad \text{(B-1)}$$

Where, $\boldsymbol{B}(t, s) = {\rm B}(t, s)\boldsymbol{e}_{||}(t, s)$ and $d\boldsymbol{A}(t, s) = \pi\, {\rm d}t^2\boldsymbol{e}_{||}(t, s)$. $d\boldsymbol{A}(t, s)$ is the *infinitesimal*, cross-sectional area of the tube, *oriented along* $\boldsymbol{e}_{||}(t, s)$ at P, with area = $\pi\, {\rm d}t^2$. Eq. (B-1) simplifies to,

$$\psi_{\rm in}(t, s) = B(t, s)\, \pi\, {\rm d}t^2 \qquad \text{(B-2)}$$

Similarly, the magnetic flux *leaving* the tube at point Q ($t$, $s$ + d$s$) is

$$\psi_{\rm out}(t, s + {\rm d}s) = \boldsymbol{B}(t, s + {\rm d}s) \cdot d\boldsymbol{A}(t, s + {\rm d}s) \quad \text{(B-3)}$$

Where, $\boldsymbol{B}(t, s + {\rm d}s) = B(t, s + {\rm d}s)\boldsymbol{e}_{||}(t, s + {\rm d}s)$, ${\rm d}\boldsymbol{A}(t, s + {\rm d}s) = \pi\, {\rm d}t^2 \boldsymbol{e}_{||}(t, s + {\rm d}s)$.
Eq. (B-3) simplifies to,

$$\psi_{\rm out}(t, s + {\rm d}s) = B(t, s + {\rm d}s)\, \pi\, {\rm d}t^2 \quad \text{(B-4)}$$

The net change in flux is

$$\Delta\psi(t, s) = \psi_{\rm out}(t, s + {\rm d}s) - \psi_{\rm in}(t, s) = [B(t, s + {\rm d}s) - B(t, s)]\pi\, {\rm d}t^2 \quad \text{(B-5)}$$

Or,

$$\Delta\psi = \left[\frac{\partial B}{\partial s} ds\right] \pi dt^2 \quad \text{(B-6)}$$

Thus, the *fractional* change in the magnetic flux is,

$$\frac{\Delta\psi}{\psi} = \frac{\left[\frac{\partial B}{\partial s} ds\right] \pi dt^2}{B(s,t)\pi dt^2} = \frac{1}{B(s,t)} \left[\frac{\partial B}{\partial s} ds\right] = \frac{\partial \ln B}{\partial s} ds \quad \text{(B-7)}$$

***Electron flux through the tube:*** Similar to the above case, one may consider the flux of electrons through the tube. The flux *entering* at P($t$, $s$) [22]

$$F_{\rm in}(t, s) = \left[\frac{1}{4} n V_{\rm the}\right](t, s)[\boldsymbol{e}_{||}(t, s)] \cdot d\boldsymbol{A}(t, s) = (\pi dt^2) \left[\frac{1}{4} n V_{\rm the}\right](t, s) \quad \text{(B-8)}$$

Likewise, the electron flux *leaving* at Q ($t$, $s$ + d$s$) is

$$F_{\rm out}(t, s + {\rm d}s) = (\pi {\rm d}t^2) \left[\frac{1}{4} n V_{\rm the}\right](t, s + {\rm d}s) \quad \text{(B-9)}$$

The net change in the electron flux at P ($t$, $s$) is

$$\Delta F = F_{\rm out}(t, s + {\rm d}s) - F_{\rm in}(t, s) = \frac{\partial}{\partial s}\left[\frac{1}{4} n V_{\rm the}\right] ds\, (\pi dt^2) \quad \text{(B-10)}$$

The *fractional* change in electron flux is given by,

$$\frac{\Delta F}{F} = \frac{\frac{\partial}{\partial s}\left[\frac{1}{4} n V_{\rm the}\right] ds (\pi dt^2)}{\left[\frac{1}{4} n V_{\rm the}\right](\pi dt^2)} = \frac{\partial}{\partial s} \ln\left[\frac{1}{4} n v_{\rm the}\right] ds \quad \text{(B-11)}$$

*Now, for electrons to be strongly bound to the field lines, the fractional change in electron flux computed in* (B-11) *above, must equal the fractional change in magnetic flux given by* (B-7). Setting these two quantities equal obtains,

$$\frac{\Delta F}{F} = \frac{\Delta\psi}{\psi} \Rightarrow \frac{\partial}{\partial s} \ln\left[\frac{1}{4} n V_{\rm the}\right] {\rm d}s = \frac{\partial \ln B}{\partial s} {\rm d}s$$
$$\Rightarrow \frac{\partial}{\partial s} \ln\left[\frac{1}{4} \frac{n V_{\rm the}}{B}\right] = 0$$
$$\Rightarrow \ln\left[\frac{1}{4} \frac{n V_{\rm the}}{B}\right] = const.$$
$$\Rightarrow \ln\left[\frac{n\sqrt{T_e}}{B}\right] = {\rm Const} \quad \text{(B-12)}$$

Where, in the last step above, $V_{\rm the} = \sqrt{\frac{8eT_e}{\pi m}}$ has been used and all constants have been absorbed in the RHS. $n/B$ scaling follows immediately from (B-12) for isothermal electrons.

**References**

[1] C. Charles, "Plasmas for spacecraft propulsion," J. Phys. D: Appl. Phys. **42**, 163001 (2009).

[2] L. Garrigues and P. Coche, "Electric propulsion: Comparisons between different concepts," Plasma Phys. Controlled Fusion **53**, 124011 (2011).

[3] E. Ahedo, "Plasmas for space propulsion," Plasma Phys. Controlled Fusion **53**, 124037 (2011).

[4] S. Shinohara, H. Nishida, T. Tanikawa, T. Hada, I. Funaki, and K. P. Shamrai, "Development of electrodeless plasma thrusters with high-density helicon plasma sources," IEEE Trans. Plasma Sci. **42**, 1245–1254 (2014).

[5] S. Mazouffre, "Electric propulsion for satellites and spacecraft: Established technologies and novel approaches," Plasma Sources Sci. Technol. **25**, 033002 (2016).

[6] D. Rafalskyi and A. Aanesland, "Brief review on plasma propulsion with neutralizer-free systems," Plasma Sources Sci. Technol. **25**, 043001 (2016).

[7] S. N. Bathgate, M. M. M. Bilek, and D. R. McKenzie, "Electrodeless plasma thrusters for spacecraft: A review," Plasma Sci. Technol. **19**, 083001 (2017).

[8] F. Ebersohn, S. Girimaji, D. Staack, J. Shebalin, B. Longmier, and C. Olsen, "Magnetic nozzle plasma plume: Review of crucial physical phenomena," in *48th AIAA/ASME/SAE/ASEE Joint Propulsion Conference & Exhibit*, AIAA 2012-4274 (2012).

[9] K. Takahashi, C. Charles, R. Boswell, and A. Ando, "Effect of magnetic and physical nozzles on

plasma thruster performance," Plasma Sources Sci. Technol. **23**, 044004 (2014).

[10] M. Merino and E. Ahedo, "Magnetic nozzles for space plasma thrusters," in *Encyclopedia of Plasma Technology*, Vol. 2, pp. 1329–1351 (CRC Press, 2017).

[11] E. Ahedo, "Using electron fluid models to analyze plasma thruster discharges," J. Electr. Propul. **2**, 2 (2023).

[12] K. Takahashi, "Helicon-type radiofrequency plasma thrusters and magnetic plasma nozzles," Rev. Mod. Plasma Phys. **3**, 3 (2019).

[13] I. D. Kaganovich *et al.*, "Physics of E × B discharges relevant to plasma propulsion and similar technologies," Phys. Plasmas **27**, 120601 (2020).

[14] E. Dale, B. Jorns, and A. Gallimore, "Future directions for electric propulsion research," Aerospace **7**, 120 (2020).

[15] K. Takahashi, "Magnetic nozzle radiofrequency plasma thruster approaching twenty percent thruster efficiency," Sci. Rep. **11**, 2768 (2021).

[16] I. Adamovich *et al.*, "The 2022 plasma roadmap: Low temperature plasma science and technology," J. Phys. D: Appl. Phys. **55**, 373001 (2022).

[17] A. Anders, "Plasma and ion sources in large area coating: A review," Surf. Coat. Technol. **200**, 1893–1906 (2005).

[18] I. Levchenko *et al.*, "Space micropropulsion systems for CubeSats and small satellites: From proximate targets to furthermost frontiers," Appl. Phys. Rev. **5**, 011104 (2018).

[19] M. Keidar, T. Zhuang, A. Shashurin, G. Teel, D. Chiu, J. Lukas, S. Haque, and L. Brieda, "Electric propulsion for small satellites," Plasma Phys. Controlled Fusion **57**, 014005 (2015).

[20] I. Levchenko, S. Xu, S. Mazouffre, D. Lev, D. Pedrini, D. Goebel, L. Garrigues, F. Taccogna, and K. Bazaka, "Perspectives, frontiers, and new horizons for plasma-based space electric propulsion," Phys. Plasmas **27**, 020601 (2020).

[21] I. Levchenko, D. Goebel, D. Pedrini, R. Albertoni, O. Baranov, I. Kronhaus, D. Lev, M. L. R. Walker, S. Xu, and K. Bazaka, "Recent innovations to advance space electric propulsion technologies," Prog. Aerosp. Sci. **152**, 100900 (2025).

[22] M. A. Lieberman and A. J. Lichtenberg, *Principles of Plasma Discharges and Materials Processing* (John Wiley & Sons, New York, 1994).

[23] G. Fridman, G. Friedman, A. Gutsol, A. B. Shekhter, V. N. Vasilets, and A. Fridman, "Applied plasma medicine," Plasma Process. Polym. **5**, 503–533 (2008).

[24] M. G. Kong, G. Kroesen, G. Morfill, T. Nosenko, T. Shimizu, J. van Dijk, and J. L. Zimmermann, "Plasma medicine: An introductory review," New J. Phys. **11**, 115012 (2009).

[25] A. Fridman and G. Friedman, *Plasma Medicine* (John Wiley & Sons, 2012).

[26] F. F. Chen, "Industrial applications of low-temperature plasma physics," Phys. Plasmas **2**, 2164–2175 (1995).

[27] R. d'Agostino, P. Favia, Y. Kawai, H. Ikegami, N. Sato, and F. Arefi-Khonsari, *Advanced Plasma Technology* (Wiley-VCH, 2008).

[28] C. G. N. Lee, K. J. Kanarik, and R. A. Gottscho, "The grand challenges of plasma etching: A manufacturing perspective," J. Phys. D: Appl. Phys. **47**, 273001 (2014).

[29] A. Fridman, *Plasma Chemistry* (Cambridge University Press, Cambridge, 2008).

[30] Y. Ju and W. Sun, "Plasma assisted combustion: Dynamics and chemistry," Prog. Energy Combust. Sci. **48**, 21–83 (2015).

[31] N. A. Popov, "Kinetics of plasma-assisted combustion: Effect of non-equilibrium excitation on the ignition and oxidation of combustible mixtures," Plasma Sources Sci. Technol. **25**, 043002 (2016).

[32] Y. Sui, C. A. Zorman, and R. M. Sankaran, "Plasmas for additive manufacturing," Plasma Process. Polym. **17**, 2000009 (2020).

[33] J. Hong, A. B. Murphy, B. Ashford, P. J. Cullen, T. Belmonte, and K. Ostrikov, "Plasma-digital nexus: Plasma nanotechnology for the digital manufacturing age," Rev. Mod. Plasma Phys. **4**, 1 (2020).

[34] S. Stauss, H. Muneoka, and K. Terashima, "Review on plasmas in extraordinary media:

Plasmas in cryogenic conditions and plasmas in supercritical fluids," Plasma Sources Sci. Technol. **27**, 023003 (2018).

[35] J. Hopwood, "Review of inductively coupled plasmas for plasma processing," Plasma Sources Sci. Technol. **1**, 109–116 (1992).

[36] A. Bogaerts, E. Neyts, R. Gijbels, and J. van der Mullen, "Gas discharge plasmas and their applications," Spectrochim. Acta Part B **57**, 609–658 (2002).

[37] R. W. Boswell, "Very efficient plasma generation by whistler waves near the lower hybrid frequency," Plasma Phys. Controlled Fusion **26**, 1147–1162 (1984).

[38] F. F. Chen, "Helicon discharges and sources: A review," Plasma Sources Sci. Technol. **24**, 014001 (2015).

[39] S. Shinohara, T. Hada, T. Motomura, K. Tanaka, T. Tanikawa, K. Toki, Y. Tanaka, and K. P. Shamrai, "Development of high-density helicon plasma sources and their applications," Phys. Plasmas **16**, 057104 (2009).

[40] A. J. Perry, D. Vender, and R. W. Boswell, "The application of the helicon source to plasma processing," J. Vac. Sci. Technol. B **9**, 310–317 (1991).

[41] H. Conrads and M. Schmidt, "Plasma generation and plasma sources," Plasma Sources Sci. Technol. **9**, 441–454 (2000).

[42] R. Geller, *Electron Cyclotron Resonance Ion Sources and ECR Plasmas* (Routledge, 2018).

[43] P. Svarnas, "Electron cyclotron resonance (ECR) plasmas: A topical review through representative results obtained over the last 60 years," J. Appl. Phys. **137**, 070701 (2025).

[44] E. Bering, M. Brukardt, J. Squire, T. Glover, V. Jacobson, and G. McCaskill, "Recent improvements in ionization costs and ion cyclotron heating efficiency in the VASIMR engine," in *44th AIAA Aerospace Sciences Meeting and Exhibit*, AIAA 2006-766 (2006).

[45] A. Ganguli, R. D. Tarey, N. Arora, R. Narayanan, and K. Akhtar, "Development of compact electron cyclotron resonance plasma source," in *2013 19th IEEE Pulsed Power Conference (PPC)*, pp. 1–5 (IEEE, 2013).

[46] A. Verma, A. Ganguli, R. Narayanan, R. D. Tarey, and D. Sahu, "Compact ECR plasma source: Its physics and application," in *Proceedings of the 4th Asia-Pacific Conference on Plasma Physics*, pp. 1–16 (2020).

[47] A. Ganguli, R. D. Tarey, N. Arora, and R. Narayanan, "Development and studies on a compact electron cyclotron resonance plasma source," Plasma Sources Sci. Technol. **25**, 025026 (2016).

[48] A. Ganguli, R. D. Tarey, R. Narayanan, and A. Verma, "Evaluation of compact ECR plasma source for thruster applications," Plasma Sources Sci. Technol. **28**, 035014 (2019).

[49] A. Verma, A. Ganguli, D. Sahu, R. Narayanan, and R. D. Tarey, "Thrust evaluation of compact ECR plasma source using 2-zone global model and plasma measurements," Plasma Sources Sci. Technol. **29**, 085007 (2020).

[50] P. Singh, A. Ganguli, R. Narayanan, and D. Sahu, "Experimental studies of hydrogen plasma produced by compact ECR plasma source," IEEE Trans. Plasma Sci. **53**, 2188–2203 (2025).